\documentclass{JHEP3}\newcommand{\format} {\JHEPformat}

\usepackage{amsmath}
\usepackage{epsfig}
\usepackage{latexsym}

\newcommand{\JHEPformat} {
\bibliographystyle{JHEP}
\newcommand{\maketitlepage} {}
\abstract{\theabstract}
\keywords{\thekeywords}
\preprint{\thepreprint}
}

\newcommand{\TITLE}[1] {\newcommand{\thetitle} {#1}\title{#1}}
\newcommand{\ABSTRACT}[1] {\newcommand{\theabstract} {#1}}
\newcommand{\AUTHOR}[1] {\newcommand{\theauthor} {#1}}
\newcommand{\ADDRESS}[1] {\newcommand{\theaddress} {#1}}
\newcommand{\DATE}[1] {\newcommand{\thedate} {#1}\date{#1}}
\newcommand{\KEYWORDS}[1] {\newcommand{\thekeywords} {#1}}
\newcommand{\PREPRINT}[1] {\newcommand{\thepreprint} {#1}}

\newcommand{\half}{{\frac{1}{2}}}

\TITLE{Analytic non-supersymmetric  background dual of a 
  confining gauge theory and the corresponding plane wave theory of Hadrons}
\AUTHOR{{\bf Stanislav Kuperstein}\footnote{\tt kupers@post.tau.ac.il
                                            \tt   cobi@post.tau.ac.il}}

\ADDRESS{School of Physics and Astronomy\\
  The Raymond and Beverly Sackler Faculty of Exact Sciences\\
  Tel Aviv University, Ramat Aviv, 69978, Israel
}

\author{
Stanislav Kuperstein and Jacob Sonnenschein\\
School of Physics and Astronomy\\
The Raymond and Beverly Sackler Faculty of Exact Sciences\\
Tel Aviv University, Ramat Aviv, 69978, Israel.\\
E-mail:
\email{kupers@post.tau.ac.il, cobi@post.tau.ac.il.}
}

\ABSTRACT{

We find a regular analytic 1st order deformation of the Klebanov-Strassler background.
From the dual gauge theory point of view  the deformation describes 
supersymmetry soft breaking gaugino mass terms.
We calculate the difference in vacuum energies between the supersymmetric
and the non-supersymmetric solutions and find that it  matches the  field theory prediction.
We also discuss  the breaking of the 
$U(1)_R$ symmetry and the space-time dependence of the gaugino bilinears two point function.
Finally, we determine the Penrose limit of the non-supersymmetric
background and write down the corresponding plane wave string
theory. This  string describes ``annulons''-heavy hadrons with mass
proportional to large global charge. Surprisingly the string spectrum 
  has two fermionic zero modes.  This  implies that the   sector in
  the  non-supersymmetric gauge theory which   is the dual of the
  annulons is supersymmetric. 
}

\DATE{August 2003}
\PREPRINT{TAUP-2746-03\\{\tt hep-th/0309011}}

\format

\begin{document}

\maketitlepage

\section{Introduction}

The AdS/CFT correspondence \cite{Maldacena:1998re},
\cite{Gubser:1998bc}, \cite{Witten:1998qj} (see
 \cite{Aharony:1999ti} for a review)
is an explicit realization of the
holography principle and describes the  duality between a
string theory (type IIB on $AdS_5 \times S^5$) and a gauge theory
($\mathcal{N}=4$ SYM in four dimensions). Since the formulation of
the AdS/CFT conjecture there has been great progress in the study of
theories with less
supersymmetries and not necessarily conformal.
There are several approaches one can use to break the $\mathcal{N}=4$
supersymmetry down to $\mathcal{N}=2$ or $\mathcal{N}=1$.
For instance, the SUSY is broken by placing D-branes at an
orbifold fixed point. One might also construct the $\mathcal{N}=2^*$
theory by giving non-zero masses to various fields in the gauge
theory. A few years ago two important examples of supergravity duals of $\mathcal{N}=1$  
gauge theories have been  provided by  \cite{Maldacena:2000yy}
and  \cite{Klebanov:2000hb}
(see \cite{Bertolini:2003iv} and \cite{Bigazzi:2003ui} for recent reviews). 
The Maldacena-Nunez (MN) background
consists of NS5-branes wrapped on an $S^2$ and based on the solution
of  \cite{Chamseddine:1997nm}. 
The supergravity dual of Klebanov-Strassler (KS) involves D5 branes wrapped around
a shrinking $S^2$. The metric has a standard D3-form with the
6d deformed conifold being the transversal part of the 10d space.

Non-supersymmetric deformations of the MN background have been studied by number of authors.
In  \cite{Aharony:2002vp} the supersymmetry was broken completely by giving masses for some of the
scalar fields. The explicit solution was constructed
in seven-dimensional gauged supergravity at leading order in the deformation
and then up-lifted to ten dimensions. It was argued that the deformed non-supersymmetric 
background is guaranteed to be stable, since the original dual gauge theory had a
mass gap. On other hand, the authors of  \cite{Evans:2002mc} used the
solution of \cite{Gubser:2001eg} to study
the supersymmetry breaking by the inclusion of a gaugino mass term and a condensate.  
Evidently, the global symmetry remains unbroken under this deformation.

Our main goal is to find a non-singular,
non-supersymmetric deformation of the KS solution,
which preserves the global symmetries of the original background and to study the Penrose limit
of the new solution. 
The problem has
been already attacked by different authors. In  \cite{Apreda:2003gc}  the second order equations
of motion following from the one dimensional effective action were
solved perturbatively in the deep IR and UV regions. However, the
numerical interpolation between the IR and UV regions does not
lead to a desired relation between the corresponding parameters and
the question of existence of the global non-singular solution remains unanswered. 
The authors of \cite{Borokhov:2002fm} suggested a computational technique for studying the
non-supersymmetric solution. The technique is based on the
modification of the first order BPS equations, so that we might continue to use 
a superpotential even for a non-supersymmetric solution. 
In short, one obtains
a set of sixteen 1st equations and one zero-order constraint
instead of eight standard 2nd order differential  equations.
Analyzing asymptotic solutions it was found that
regularity of the IR and UV behavior restricts to three the number of possible deformations.

In this paper we determine and describe  a regular \emph{analytic} solution of the 1st order equations 
similar to those appearing in \cite{Borokhov:2002fm}. 
We note that these equations are significantly 
simplified once we properly redefine the radial coordinate.
(The equations transform non-trivially under the coordinate redefinition since
one has to apply the ``zero-energy'' constraint, which removes the ``gauge freedom'' of the
coordinate transformation).
We also demonstrate how part of the 1st order equations can be re-derived using 
the usual 2nd order IIB equations of motion.

Our solution preserves the global symmetry and therefore
describes a deformation corresponding to the inclusion of mass terms
of the two gaugino bilinears in the dual gauge theory.

Taking Penrose limits of both MN and KS solutions around geodesics located in the 
deep IR region results in solvable string theory models\cite{Gimon:2002nr}.
The string Hamiltonians of these models describe the 3d non-relativistic motion
and the excitations of heavy hadrons (called ``annulons'') with masses proportional to a large global 
symmetry charge $M_{annulon}=m_0 J$.
It was further shown that the $\mathcal{N}=1$ supersymmetry of the original theories implies 
that the world-sheet spinors have two zero-frequency modes providing 
a 4-dimensional Hilbert sub-space of degenerate states (two bosonic and two fermionic).

We construct a Penrose limit (see \cite{Gueven:2000ru}, \cite{Blau:2001ne},
\cite{Metsaev:2001bj}, \cite{Berenstein:2002jq},
\cite{Metsaev:2002re} and \cite{Blau:2002mw})
of our non-super\-sym\-met\-ric 
KS background and obtain a pp-wave metric
and a complex 3-form which are very  similar to the PL limit \cite{Gimon:2002nr}
of the supersymmetric solution.
We also quantize the light-cone string Hamiltonian and determine the
masses  of the bosonic and fermionic modes. These masses, though
different from the supersymmetric case, still obey the relation that
the sum of the mass squared is the same for bosonic and fermionic
modes.
Again the  string describes kinematics and excitations of the annulons. The only difference between them
and those of \cite{Gimon:2002nr} is a modification of $m_0$.
A  surprising feature of the string spectrum is that, like in the
Penrose limit of the KS background, here as well, there are 
two fermionic zero modes.
In the dual field theory this implies that even though the full theory is non-supersymmetric,
the sector of states with large $J$ charge  admits supersymmetry.
It is conceivable that  in this limit of states of large mass the 
impact of the small  gaugino mass deformations is washed away.

The authors of \cite{Apreda:2003gs}
used the  solution of \cite{Apreda:2003gc} to take the PL.
The IR expansion of the fields given in \cite{Apreda:2003gc} differs, however,
from our solution (see later) and therefore the pp-wave background
of \cite{Apreda:2003gs} is also slightly different from the metric we have
derived.

The organization of the paper is as follows. 
In Section \ref{section:TKSmab} we give a short review of the KS model.
Section \ref{section:NseoKS} 
is devoted to the non-supersymmetric deformation. We start by recalling the technique of 
\cite{Borokhov:2002fm}
and then derive and solve a set of 1st order equations 
using a different choice of the radial coordinate.
Since the expression of the various fields in the solution are too complicated we 
report here only the asymptotic behavior of the fields in the UV and in the IR. 
In Section \ref{section:Ve} we find the vacuum energy of the supergravity solution.
We further use this result in Section \ref{section:Dgt} while discussing various properties of 
the gauge theory dual of the non-supersymmetric background.
We argue that the deformation corresponds to the mass terms
of the two gaugino bilinears that in particular  lift the degeneracy
of the vacuum.
In Section  \ref{section:Tpwl} 
we investigate the Penrose limit of the non-supersymmetric
background. We then describe the solution of the plane wave string in
\ref{section:TplwtatA}
in the form of the annulons. 
We close in the last section with conclusions and
suggestions for a further research.  The type IIB equations of motion
and the explicit solution for some of the field are presented in two appendices

\section{The Klebanov-Strassler model and beyond}
\label{section:TKSmab}

Before reviewing the main features of the KS solution it will be worth to
write down  the type IIB equation of motion for a case of  
a constant dilaton ($e^\Phi =g_s$), a vanishing
axion ($C_0=0$) and with the 10d metric
and the 5-form flux having the structure of the D3-brane solution, namely:
\begin{equation}              \label{eq:metric}
ds^2 = h^{-1/2} \left( \mathbf{d} x_0^2 + \ldots + \mathbf{d} x_3^2 \right) +
       h^{1/2} ds^2_{M_6}
\end{equation}
and
\begin{equation}              \label{eq:5form}
\tilde{F}_5 = \frac{1}{g_s} (1 + \star_{10}) \mathbf{d} h^{-1} \wedge \mathbf{d} x_0 \wedge \ldots
                               \wedge \mathbf{d} x_3,
\end{equation}
where $M_6$ is a 6d Ricci flat transversal space and the harmonic
function $h$ depends only on the coordinates on $M_6$.
We will denote the Hodge dual on  $\mathnormal{M}_6$ by $\star_6$.
For
a general $n \leq 6$ form on $\mathnormal{M}_6$  there
is a simple relation between $\star_{10}$ and  $\star_6$:
\begin{eqnarray} 
\star_{10} \omega_n = h^{-\frac{n-1}{2}} \star_6 \omega_n
               \wedge \mathbf{d} x_0 \wedge\mathbf{d} x_1 \wedge\mathbf{d} x_2 \wedge\mathbf{d} x_3.
\end{eqnarray}
With this observation in mind and assuming that the 3-forms have their
legs only along $M_6$ we derive the 6d version of the dilaton and
axion equations of motion:
\begin{eqnarray}                 \label{eq:PhiC}
0  &=& 
      -  H_3 \wedge \star_6 H_3 + 
        g_s^2 F_3 \wedge \star_6 F_3          \nonumber\\
0  &=&  H_3 \wedge \star_6 F_3. 
\end{eqnarray}
In order to find the connection between the 3-forms and the warp
function $h$ we have to use the 5-form equation. We end up with:
\begin{equation} \label{eq:dsdh}
\mathbf{d} \star_6 \mathbf{d} h = g_s H_3 \wedge F_3
\end{equation}
or
\begin{equation}
\tilde{F}_5 =  B_2 \wedge F_3 + \star_{10} \left( B_2 \wedge F_3 \right).
\end{equation}
In what follows we will adopt the integrated version of
(\ref{eq:dsdh}):
\begin{equation} \label{eq:dh}
 \mathbf{d} h = - g_s \star_6 \left( B_2 \wedge F_3 \right).
\end{equation}
On equal footing we might replace  (\ref{eq:dh}) by
\begin{equation} \label{eq:xdh}
 \mathbf{d} h =  g_s \star_6 \left( C_2 \wedge H_3 \right).
\end{equation}
Note that the difference between the warp functions
satisfying (\ref{eq:dh}) and (\ref{eq:xdh}) is precisely the harmonic
function $\tilde{h}(\tau)$
on the deformed conifold satisfying $\bar{\nabla}_2^6 \tilde{h}=0$. 
This function, however, diverges at $\tau \to 0$. We can therefore use
one of the integrated versions of (\ref{eq:dsdh}) together with the
requirement of regularity of $h(\tau)$ at $\tau=0$.

Next we consider the 3-forms equations. Applying (\ref{eq:dh}) and the relation between 
$F_5$ and $\tilde{F}_5$ we get:
\begin{eqnarray}           \label{eq:F3H3}
\mathbf{d} \left[ h^{-1} \left( \star_6  F_3  +  \frac{1}{g_s}  H_3  \right) \right] =0
\quad
\textrm{and}
\quad
\mathbf{d} \left[ h^{-1} \left( \star_6  H_3  -  g_s F_3  \right) \right] =0
\end{eqnarray}
In deriving this result we have used the fact that
all the forms have their legs along the 6d space and therefore 
$\left(  C_2 \wedge H_3 + B_2 \wedge F_3 \right) \wedge H_3=0$.
Finally, we re-write the metric equation of motion. Remarkably, for 
the metric and the forms considered in our case it is enough to verify
only the trace of the Einstein equation. 
Calculating the Ricci scalar of the metric (\ref{eq:metric}) we find:
\begin{eqnarray}    
R = \half h^{-3/2} \nabla_6^2 h = \half h^{-3/2} \star_6 \mathbf{d} \star_6 \mathbf{d} h .
\end{eqnarray}
Recalling the self-duality of
$\tilde{F}_5$ and using again the relation between $\star_{10}$ and
$\star_6$ we obtain:
\begin{eqnarray}        \label{eq:Rmn}
\mathbf{d} \star_6 \mathbf{d}  h  &=& \frac{1}{2} \left[ H_3 \wedge \star_6 H_3 
                                +  g_s^2 F_3 \wedge \star_6 F_3 \right]. 
\end{eqnarray}
The equations we have written (\ref{eq:PhiC},\ref{eq:dh},\ref{eq:F3H3},\ref{eq:Rmn}) 
are easily solved by
requiring that:
\begin{eqnarray}        \label{eq:dualFH}
 \star_6  F_3  =  - g_s^{-1}  H_3
\quad
\textrm{and}
\quad
 \star_6  H_3  =  g_s  F_3.
\end{eqnarray}
In this case the complex form $G_3 \equiv F_3 + \frac{i}{g_s} H_3$ is
imaginary self dual $ \star_6 G_3 =i G_3$.

Note that
the equation for $h$ is a first order differential equation, even
though the solution is not supersymmetric in general.

The most important example of the supersymmetric solution is the
Klebanov-Strassler model \cite{Klebanov:2000hb}, where the 6d manifold is the deformed
conifold space. The 6d metric is given by:
\begin{eqnarray}     \label{eq:ds6}
ds_6^2 = \half \epsilon^{4/3} K(\tau) \Bigg[ 
  \frac{1}{K^3(\tau)} \left( d \tau^2 + (g^5)^2 \right) +
  \cosh^2 \left( \frac{\tau}{2} \right) \left((g^3)^2 + (g^4)^2 \right)  \nonumber\\
  +  \sinh^2 \left( \frac{\tau}{2} \right) \left((g^1)^2 + (g^2)^2 \right)
 \Bigg],
\end{eqnarray}
where
\begin{equation}
K(\tau) = 2^{-1/3} \frac{(\sinh 2 \tau - 2 \tau)^{1/3}}{\sinh \tau}.
\end{equation}
At large $\tau$ we may use another radial coordinate defined by:
\begin{equation}
r \to \epsilon^{2/3} e^{\tau/3}.
\end{equation} 
In terms of $r$ we have:
\begin{equation}
ds_6^2  \to  dr^2 + r^2 d \Omega^2_{T_{1,1}}.
\end{equation}
The determinant of (\ref{eq:ds6}) vanishes at $\tau=0$ reflecting 
the fact the 6d metric degenerates into the metric of $S^3$.

The $\mathbf{M}$ fractional D5-branes wrapping the shrinking $S^2$ are introduced through the RR 3-form:
\begin{eqnarray}     \label{eq:F3}
F_{3} = \mathbf{M} \Big[ (1-F(\tau)) g^5 \wedge g^3 \wedge g^4 +
                     F(\tau) g^5 \wedge g^1 \wedge g^2     \nonumber \\
                     +F^\prime(\tau) d\tau \wedge \left( g^1 \wedge g^3 + g^2 \wedge g^4 \right)
          \Big],
\end{eqnarray}
together with the boundary conditions $F(0)=0$ and $F(\infty)=\half$.
The former condition ensures
that $F_3$ is proportional to the volume form of the non-collapsing $S^3$ at $\tau=0$,
while the later means the restoration of the $U(1)_R$ symmetry in the UV, 
where we approach the geometry of the
conifold over $T_{1,1}$ \cite{Klebanov:2000hb}. On using the duality relations (\ref{eq:dualFH}) one may find
the NS 3-form:
\begin{eqnarray}     \label{eq:H3}
H_{3} = \mathbf{d} B_2 = g_s \mathbf{M} \Big[ 
                     f^\prime(\tau) d \tau \wedge g^1 \wedge g^2 +
                     k^\prime(\tau) d \tau \wedge g^3 \wedge g^4 \nonumber \\
                     +\half \left( k(\tau)-f(\tau) \right) 
                        g^5 \wedge \left( g^1 \wedge g^3 + g^2 \wedge g^4 \right)
          \Big],
\end{eqnarray}
where the functions $f(\tau)$,$k(\tau)$ and $F(\tau)$ satisfy:
\begin{eqnarray}          \label{eq:fkF}
f^\prime(\tau) = (1-F(\tau)) \tanh^2 \left( \frac{\tau}{2} \right),
\quad
k^\prime(\tau) = F(\tau) \coth^2 \left( \frac{\tau}{2} \right),
\nonumber \\
\textrm{and} \quad
F^\prime(\tau) = \frac{k(\tau)-f(\tau)}{2}. \qquad \qquad \qquad 
\end{eqnarray}
This system of the first order differential equations 
has three dimensional space of solutions:
\begin{eqnarray}     \label{eq:fkFsol}
F(\tau) &=& \half - \frac{\tau}{2 \sinh \tau} +
        C_1 \left( \cosh \tau - \frac{\tau}{\sinh \tau} \right) +
        \frac{C_2}{\sinh \tau}
\nonumber \\
f(\tau) &=& \frac{\tau \coth \tau -1}{2 \sinh \tau}(\cosh \tau - 1) +
\nonumber \\           
&&        + C_1 \left(2 \tau - \sinh \tau - \tanh \frac{\tau}{2} - 
                              \frac{\tau}{2 \cosh^2 \frac{\tau}{2}} \right) +
        \frac{C_2}{2 \cosh^2 \frac{\tau}{2}}  + C_3
\nonumber \\    
k(\tau) &=& \frac{\tau \coth \tau -1}{2 \sinh \tau}(\cosh \tau + 1) +
\nonumber \\           
&&        + C_1 \left(2 \tau + \sinh \tau - \coth \frac{\tau}{2} + 
                              \frac{\tau}{2 \sinh^2 \frac{\tau}{2}} \right) -
        \frac{C_2}{2 \sinh^2 \frac{\tau}{2}}  + C_3.
\end{eqnarray}
The KS solution corresponds to $C_1=C_2=C_3=0$. Using the complex structure of 
the deformed conifold space
the complex form $G_{3}= F_{3}+ \frac{i}{g_s} H_{3}$
can be identified in this case as a $(2,1)$ form \cite{Herzog:2001xk}.
Instead, for $C_1=-\half$ and  $C_2=C_3=0$
we obtain a $(0,3)$ form which breaks the supersymmetry and diverges
at $\tau \to \infty$.
In Appendix \ref{section:03} we prove this statement by performing an  explicit
calculation making use of the complex structure of the deformed
conifold given in  \cite{Papadopoulos:2000gj} (similar derivation can be done
using the results of \cite{Cvetic:2000db} and \cite{Herzog:2001xk}).
\footnote{Note that the deformed conifold is a non-compact space
and therefore there is no obstacle for constructing a closed $(0,3)$
form, which is regular everywhere except at infinity.}
For  $C_2 \neq 0$ we find another  $(2,1)$ form which is singular at $\tau=0$.
Finally, the solution with $C_3\neq 0$ amounts to the gauge freedom
$B_2 \to B_2 + \mathbf{d} A_1$.

Having determined the 3-forms one can integrate (\ref{eq:dh}) 
to find the harmonic function $h(\tau)$:
\begin{equation}
h(\tau)= \alpha \frac{2^{1/3}}{4} \int_{\tau}^{\infty} dx \frac{x \coth x -1}{\sinh^2 x}
                                                         (\sinh(2x)-2x)^{1/3}.
\end{equation}
The asymptotic behavior of $h(\tau)$ is:
\begin{equation}             \label{eq:hASYMP}
h(\tau \to 0) \to a_0 \quad \textrm{and} \quad  
 h(\tau \to \infty) \to  \alpha \frac{3}{4} 2^{1/3} \tau e^{-4\tau/3}
\end{equation}
with $\alpha \equiv 4 \left(g_s \mathbf{M}l_s^2 \right)^2 \epsilon^{-8/3}$.

The dual field theory realized on the world-volume of the $\mathbf{N}$ physical and $\mathbf{M}$ 
fractional D3-branes is a 4d $\mathcal{N}=1$ supersymmetric 
$SU(\mathbf{N}+\mathbf{M}) \times SU(\mathbf{N})$ gauge theory with a $SU(2) \times SU(2)$
global symmetry inherited from the conifold isometries.
The gauge theory is coupled to two bi-fundamental
chiral multiplets $A$ and $B$, which transform as a doublet of one of
the $SU(2)$'s each and are inert under the second $SU(2)$.
This theory is believed to exhibit a cascade of Seiberg dualities
reducing in the deep IR to pure $SU(\mathbf{M})$.
On the supergravity side $\mathbf{M}$ is fixed by the charge of the RR 3-form,
while $\mathbf{N}$ is encoded in the UV behavior of the 5-form:
\begin{equation}
\mathcal{F}_5 \sim \mathbf{N}_{\textrm{eff}} \textrm{Vol} (T_{1,1}),
\quad \textrm{where} \quad
 \mathbf{N}_{\textrm{eff}} = \mathbf{N} + \frac{3}{2 \pi} g_s \mathbf{M}^2 \ln{\frac{r}{r_0}}.
\end{equation}
The sum of the gauging couplings is constant and the logarithmic running of
the difference is determined by the NS 2-form:
\begin{equation}
\frac{1}{g_1^2} + \frac{1}{g_2^2} \sim e^{-\Phi}
\qquad
\frac{1}{g_1^2} - \frac{1}{g_2^2} \sim e^{-\Phi} \left[ \int_{S^2} B_2 -\frac{1}{2} \right].
\end{equation} 
Similarly to pure $SU(\mathbf{M})$ the theory confines. This is evident by virtue of the fact that the warp
factor approaches a constant value $h_0 \sim a_0$ at $\tau \to 0$ and therefore the tension of the confining strings
does not diverge. This conclusion is valid only for a non-zero value of the deformation parameter $\epsilon$,
since $a_0 \sim \epsilon^{-8/3}$. 
Note also that for $\epsilon \neq 0$ the $U(1)_R$ conifold symmetry is
broken down to $\mathbf{Z}_2$. This is the symmetry preserved by the gaugino bilinear
$\textrm{Tr} \lambda\lambda(x)$.
In the supergravity dual this gauge theory operator is associated with the form 
$C_2 = C^{RR}_2 + i B^{NS}_2$ \cite{Loewy:2001pq}. Subtracting the asymptotic value of $G_3=\mathbf{d} C_2$
we find at $\tau \to \infty$:
\begin{equation}     \label{eq:DG}
\Delta G_3 \approx \half \mathbf{M} \tau e^{-\tau} \omega_3, \qquad 
\omega_3 =  g^5 \wedge\left[ \left( g^3 \wedge g^4 - g^1 \wedge g^2 \right) +
                  i g_s \left( g^1 \wedge g^3 + g^2 \wedge g^4 \right)  
           \right],     
\end{equation}
where we write only the polarization along $T_{1,1}$.
Similarly:
\begin{equation}       \label{eq:DC}
\Delta C_2 \approx - \half \mathbf{M} \tau e^{-\tau} \omega_2, \qquad
\omega_2 = \left[ \left( g^1 \wedge g^3 + g^2 \wedge g^4 \right) +
                  i g_s \left( g^1 \wedge g^2 - g^3 \wedge g^4 \right)  
           \right]     
\end{equation}
and we see that $\Delta C_2$ transforms under $U(1)_R$ by the same phase as 
$\textrm{Tr} (\lambda\lambda)$. Moreover, $\Delta G_3$ has an asymptotic behavior 
we would expect from a scalar operator of dimension 3 and a non-zero VEV, namely:
\begin{equation}
\Delta G_3 = \half \mathbf{M} \frac{m^3}{r^3} \ln \frac{r^3}{m^3} \omega_3,
\end{equation}
where the deformation parameter is related to the 4d mass scale through $m \sim \epsilon^{2/3}$.

Finally, we will recall the identification of supergravity fields with   
gauge theory operators. In order to find this correspondence one writes the most
general $SU(2) \times SU(2)$ invariant background ansatz, which includes the supersymmetric
KS solution:
\begin{eqnarray}
ds^2 &=& 2^{1/2} 3^{3/4} \Big[ 
   e^{-5q(\tau)+2 Y(\tau)} \left( d x_{\mu}d x^{\mu} \right) + 
   + \frac{1}{9} e^{3q(\tau)-8p(\tau)} \left( d\tau^2 + g_5^2 \right) +  \nonumber\\
&& \quad +  \frac{1}{6} e^{3q(\tau)+2p(\tau)+y(\tau)} \left( g_1^2 + g_2^2 \right) +
             \frac{1}{6} e^{3q(\tau)+2p(\tau)-y(\tau)} \left( g_3^2 + g_4^2 \right)
\Big] 
 \nonumber \\
B_2 &=& - \left( \tilde{f}(\tau) g_1 \wedge g_2 + \tilde{g}(\tau) g_3 \wedge g_4 \right),
\quad \Phi= \Phi(\tau),
 \nonumber \\
F_3 &=& 2 P  g_5 \wedge g_3 \wedge g_4 + \mathbf{d}\left[\tilde{F}(\tau) \left( g_1 \wedge g_3 + g_2 \wedge g_4 \right)
                                          \right]  ,
 \nonumber \\
\tilde{F}_5 &=& \mathcal{F}_5 + \star_{10} \mathcal{F}_5, \quad
   \mathcal{F}_5 = - \tilde{L}(\tau)  g_1 \wedge g_2 \wedge g_3 \wedge g_4 \wedge g_5,
 \nonumber \\
 && \tilde{L}(\tau) = Q + \tilde{f}(\tau) (2P-\tilde{F}(\tau)) +  \tilde{k}(\tau) \tilde{F}(\tau).
\end{eqnarray}
This general ansatz includes both the conformal solution with a singular geometry
 ($y=\tilde{f}-\tilde{k}=0$) and the non-conformal case with regular deformed conifold 
($y,\tilde{f}-\tilde{k} \neq 0$). Here $\tilde{f}, \tilde{k}$ and $\tilde{F}$  
are the rescaled KS functions:
\begin{equation}
\tilde{f} = - 2 P g_s f , \qquad \tilde{k} = - 2 P g_s k, \qquad 
\tilde{F} =  2 P F
\end{equation}
and the constants $Q$ and $P$ are related to the number of physical and fractional branes
respectively: $P= \frac{1}{4} \mathbf{M}l_s^2$ and $Q$ is proportional to $\mathbf{N}$,
but for $P \neq 0$, it  can be re-absorbed
in the redefinition of  $\tilde{f}$ and $\tilde{k}$ \footnote{In what follows we will use the $Q=0$ convention}.
Note that for the given structure of the 3-form $F_3$ the integral $\int_{S_{3}} F_3$
does not depend on $\tilde{F}(\tau)$. Moreover, the NS-NS 3-form has the same structure
as in the KS solution as dictated by the equation for a vanishing axion $H_3 \wedge F_3=0$.

In the next section we will also use another parameterization of the 10d metric:
\begin{eqnarray}         \label{eq:6dmn}
ds^2 &=& 
        h^{-1/2} \left( d x_{\mu}d x^{\mu} \right) + 
        h^{1/2} \Big( e^{m-\frac{n}{2}}  \left( d\tau^2 + g_5^2 \right) +
\nonumber \\
 && \quad
                     + e^{\frac{n}{2}+y}  \left( g_1^2 + g_2^2 \right)  +
                       e^{\frac{n}{2}-y}  \left( g_3^2 + g_4^2 \right)  
                \Big). 
\end{eqnarray}
In particular in the supersymmetric case we have:
\begin{equation}
e^{m_0(\tau)}= \frac{\epsilon^{8/3}}{24} K(\tau)^{-1} \sinh \tau, \quad
e^{n_0(\tau)}= \frac{\epsilon^{8/3}}{16} K^2(\tau) \sinh^2 \tau, \quad
e^{y_0(\tau)}= \tanh \left( \frac{\tau}{2} \right)
\end{equation}
The two parameterizations are connected by:
\begin{equation}        \label{eq:mnhYpq}
e^{10q(\tau)-4Y(\tau)} = 3^{3/2} 2 h(\tau) , \quad
e^{4p(\tau)+4Y(\tau)-4q(\tau)} = \frac{1}{3} e^n(\tau) , \quad
e^{-10p(\tau)} = \frac{3}{2} e^{m(\tau)-n(\tau)}.
\end{equation}
Integration of the type IIB Lagrangian over the angular and the world-volume coordinates 
yields a 1d effective action
\cite{Klebanov:2000nc},\cite{PandoZayas:2000sq},\cite{Papadopoulos:2000gj}, \cite{Bigazzi:2000uu}:
\begin{eqnarray}
S \sim \int d \tau \left( - \half G_{ij} \dot{\phi}^i \dot{\phi}^j - V(\phi) \right),     
\end{eqnarray}
where \footnote{Here we adopt the conventions of \cite{Bigazzi:2003ui}.}
\begin{eqnarray}
      G_{ij} \dot{\phi}^i \dot{\phi}^j &=&
 e^{4p-4q+4Y} \Big( - 18 \dot{Y}^2 + 45 \dot{q}^2 + 30 \dot{p}^2 
   + \frac{3}{2} \dot{y}^2 + \frac{3}{4} \dot{\Phi}^2 + 
\nonumber \\
  &&   + 3 e^{-\Phi -6q-4p} \left( \frac{\sqrt{3}}{2} e^{-2y} \dot{{\tilde{f}}}^2 +
                               \frac{\sqrt{3}}{2} e^{2y} \dot{{\tilde{k}}}^2 
                        \right) +
    3 \sqrt{3} e^{\Phi -6q-4p} \dot{{\tilde{F}}}^2   \Big)
\nonumber \\
    V(\phi) &=& e^{4Y} \Bigg( \frac{1}{3} e^{-16p-4q} - 2 e^{-6p-4q} \cosh y +
                  \frac{3}{4} e^{4p-4q} \sinh^2 y + 
\nonumber \\
    &&          + \frac{3 \sqrt{3}}{4} e^{\Phi-10q+2y} (2P-\tilde{F})^2 +
                  \frac{3 \sqrt{3}}{4} e^{\Phi-10q-2y} \tilde{F}^2 +
\nonumber \\
    &&          + \frac{3 \sqrt{3}}{8} e^{-\Phi-10q} (\tilde{k}-\tilde{f})^2 +
                  \frac{9}{2} e^{-4p-16q} \tilde{L}^2
                        \Bigg) .
\end{eqnarray}
There is also a ``zero-energy'' constraint:
\begin{equation}  
 \half G_{ij} \dot{\phi}^i \dot{\phi}^j - V(\phi) = 0
\end{equation}
This Lagrangian admits a superpotential
\begin{eqnarray}
&  \displaystyle{
  V = \frac{1}{8}  G^{ij}\frac{\partial W}{\partial \phi^i} \frac{\partial W}{\partial \phi^j} 
}&
\nonumber \\
&\textrm{for} \qquad
    W= - 3 e^{4Y+4p-4q} \cosh y - 2 e^{4Y-6p-4q} - 3 \sqrt{3} e^{4Y-10q} \tilde{L} &
\end{eqnarray}
and for supersymmeric solutions the second order equations of motion
can be reduced to the first order ones:
\begin{equation}
\frac{d \phi^i}{d \tau} = \half G^{ij} \frac{\partial W}{\partial \phi^j}.
\end{equation}
 The potential appearing in the action 
has an $\mathcal{N}=1$ critical point corresponding to the conformal background
$AdS_5 \times T_{1,1}$ generated by physical D3-branes in absence of fractional branes ($P=0$).
Expanding the potential around the critical point and using the mass/ dimension formula
$\Delta = 2 + \sqrt{4 +m^2}$ one obtains the dimensions of the fields,
which now can be identified with  various gauge theory operators
\cite{Ceresole:1999zs}, \cite{Bigazzi:2000uu}. 
Here we list two of them:
\begin{eqnarray}  \label{eq:correspondence}
y     &\to& \textrm{Tr} \left( W_{(1)}^2 - W_{(2)}^2 \right) \qquad  \Delta=3,
\nonumber \\
\xi_2 \sim -F+\frac{k-f}{2} &\to& \textrm{Tr} \left( W_{(1)}^2+W_{(2)}^2 \right)  \qquad \Delta=3.
\end{eqnarray}
There are also two massless fields. $s=f+k$ is associated with a marginal direction in the CFT 
and the corresponding operator is $\textrm{Tr}\left(F_{(1)}^2-F_{(2)}^2\right)$. Similarly, 
the dilaton $\Phi$ corresponds to 
$\textrm{Tr}\left(F_{(1)}^2+F_{(2)}^2\right)$.

In this paper we will focus on the non-supersymmetric deformation
of the KS background by introducing mass terms of the gaugino bilinears 
associated with both $\xi_2$ and $y$. The former field is related to the 
SUGRA 3-forms and the latter is responsible for a deformation of the 6d metric.
The expected UV behavior of the fields in the background deformed by the masses is
$g(\tau) e^{-\tau/3}$, where $g(\tau)$ is a polynomial in $\tau$.

To conclude this section let us add a remark supporting the correspondence
(\ref{eq:correspondence}). As we have already discussed inspecting the
UV behavior of the 3-form $G_3$ one can identify it with a gaugino
bilinear in the gauge theory. This observation is related to the second line 
of (\ref{eq:correspondence}). To justify the first
line in a similar way  let us expand the 10d metric  (\ref{eq:6dmn}) at $\tau \to \infty$.
Keeping only the parts including $g_1,\ldots,g_4$ (note that $g_5$ is
invariant under $U(1)_R$) and omitting the overall factor we get:
\begin{equation}
g_1^2+g_2^2+g_3^2+g_4^2+ 2e^{-\tau} \left(g_1^2 - g_3^2 + g_2^2 - g_4^2  \right), 
\end{equation}
where we used the expansion $e^{y_0} \sim 1 -
2e^{-\tau}+\ldots$. Thus, much like the $(0,3)$ form case, 
the sub-leading term of the 6d metric transforms
under $U(1)_R$ similarly to $\xi_2$ and hence breaks the $U(1)_R$ symmetry.
Moreover it has the dimension  
of the supergravity dual of the gaugino bilinear
matching the relation in (\ref{eq:correspondence}).

\section{Non-supersymmetric extension of KS}
\label{section:NseoKS}

We start this section with a brief review of the method proposed by
\cite{Borokhov:2002fm}
(see also \cite{Ferrara:1995ih}, \cite{Freedman:1999gp},
\cite{Skenderis:1999mm}, \cite{DeWolfe:1999cp}, \cite{deBoer:1999xf} and \cite{Wijnholt:1999vk})
to study  first order non-supersymmetric deformations of the KS background
still making use of the superpotential.
We expand the fields around a given supersymmetric solution derived from the superpotential:
\begin{equation}        \label{eq:expansion}
\phi^i = \phi^i_0 + \delta \cdot \bar{\phi}^i + O(\delta^2).
\end{equation}
Define new functions:
\begin{equation}
\xi_i = G_{ij}(\phi_0) \left( \frac{d \bar{\phi}^j}{d \tau} - M^j_k(\phi_0) \bar{\phi}^k \right)
\quad
\textrm{where}
\quad
M^j_k = \half \frac{\partial}{\partial \phi^k} \left( G^{jl} \frac{\partial W}{d \phi^l} \right) . 
\end{equation}
Now one might represent the linearized equations of motion as a ``double'' set 
of first order equations:
\begin{eqnarray}   \label{eq:1st}
\frac{d \xi_i}{d \tau} + \xi_j M^j_i (\phi_0) &=& 0
\nonumber \\
\frac{d \bar{\phi}^i}{d \tau} - M^i_j (\phi_0) \bar{\phi}^j &=& G^{ik} (\phi_0) \xi_k.
\end{eqnarray} 
The second line follows trivially from the definition of $\xi_i$, while the first one 
is demonstrated by substituting the expansion (\ref{eq:expansion}) into the equations of motion 
(we refer the reader to \cite{Borokhov:2002fm} for the proof). 
Finally, the zero-energy condition can be rephrased as:
\begin{equation}
\xi_k \frac{d \bar{\phi}^k}{d \tau} = 0.
\end{equation}
An important remark is in order. One can use various definitions 
for the radial coordinate in the 1d effective action. This ambiguity is removed by applying the 
zero-energy constraint. 
The explicit form of the 1st order equations (\ref{eq:1st}) is highly dependent on the radial coordinate choice.
In our paper we will fix this ``gauge freedom'' by requiring that even in the deformed solution the
$G_{\tau\tau}$ and $G_{55}$ entries of the metric 
will remain equal exactly as in the supersymmetric case.
We will see that with this choice the set of the equations (\ref{eq:1st}) possesses an analytic solution.
On the contrary the radial coordinate ($\tau_\star$) of \cite{Borokhov:2002fm} is related to our coordinate ($\tau$) 
via $d \tau_\star = e^{4 \bar{p} - 4 \bar{q}} d \tau$. Note, however, that 
since both $\bar{p}(\tau)$ and
$\bar{q}(\tau)$ are expected to vanish at $\tau \to 0$ and $\tau \to \infty$, the 
deep UV and IR expansions of the fields have to be the same in terms of  $\tau$ and $\tau_\star$.

Let us first write the equations of motion for $\xi_i$'s \footnote{We will set $g_s=1$
throughout this section}:
\begin{eqnarray}   \label{eq:}
\dot{\xi}_{Y} &=& 0
\nonumber \\
\dot{\xi}_{q} &=& 2 \sqrt{3} e^{-4p_0 -6q_0} \tilde{L}_0 \left( \xi_Y + \xi_q \right) 
\nonumber \\
\dot{\xi}_{p} &=& \frac{4}{3} \sqrt{3} e^{-4p_0 -6q_0} \tilde{L}_0 \left( \xi_Y + \xi_q \right) 
                 + e^{-10p_0} \left( \frac{20}{9} \xi_Y + \frac{8}{9} \xi_q + 2 \xi_p \right)
\nonumber \\
\dot{\xi}_{y} &=&  - \left( \frac{1}{3} \xi_Y + \frac{2}{15} \xi_q - \frac{1}{5} \xi_p \right) \sinh y_0
                    + \xi_y \cosh y_0 +
\nonumber \\ 
&& \quad    + 2  e^{2 y_0} (2P -\tilde{F}_0) \xi_{\tilde{f}} 
                    - 2  e^{-2 y_0} \tilde{F}_0 \xi_{\tilde{k}}
\nonumber \\
\dot{\xi}_{\tilde{f}+\tilde{k}} &=& -\frac{\sqrt{3}}{3} 2 P e^{-4p_0 -6q_0}  \left( \xi_Y + \xi_q \right) 
\nonumber \\
\dot{\xi}_{\tilde{f}-\tilde{k}} &=& - \xi_{\tilde{F}}
              - \frac{2 \sqrt{3}}{3} (P -\tilde{F}_0) e^{-4p_0 -6q_0}  \left( \xi_Y + \xi_q \right) 
\nonumber \\
\dot{\xi}_{\tilde{F}} &=& 
       - \left( \cosh(2y_0) \xi_{\tilde{f}-\tilde{k}} + \sinh(2y_0) \xi_{\tilde{f}+\tilde{k}}  \right)
              - \frac{\sqrt{3}}{3} (\tilde{k}_0 -\tilde{f}_0) e^{-4p_0 -6q_0}  \left( \xi_Y + \xi_q \right) 
\nonumber \\
\dot{\xi}_{\Phi} &=&   \left( e^{2y_0} (2P-\tilde{F}_0) \xi_{\tilde{f}} 
                                + e^{-2y_0} \tilde{F}_0 \xi_{\tilde{k}}     \right)
                    -  \frac{\tilde{k}_{0}-\tilde{f}_{0}}{2} \xi_{\tilde{F}},
\end{eqnarray} 
where $\xi_{\tilde{f} \pm \tilde{k}} = \xi_{\tilde{f}} \pm \xi_{\tilde{k}}$. Throughout this paper we will be
interested in a solution satisfying:
\begin{equation}
\xi_{Y}=\xi_{p}=\xi_{q}=0.
\end{equation}
Under this assumption we have $\xi_{\tilde{f}+\tilde{k}}=X$ for
constant $X$ and from the equations for 
$\xi_{\tilde{f}-\tilde{k}}$ and $\xi_{\tilde{F}}$ we obtain:

\begin{equation}
\frac{d^2 \xi_{\tilde{f}-\tilde{k}}}{d \tau^2} = \half \left( e^{2y_0} \left(\xi_{\tilde{f}-\tilde{k}} + X \right)
                                                         +   e^{-2y_0} \left(\xi_{\tilde{f}-\tilde{k}} - X \right)
                                                       \right)
\end{equation}
This equation has a two dimensional space of solutions. However, solving for $\xi_y$,
plugging the result into the zero-energy constraint $\xi_i \dot{\phi}_0^i =0$ and requiring also regularity at
$\tau \to 0$ (otherwise we might obtain a singular solution for the fields) 
we pick up a unique simple solution $ \xi_{\tilde{f}-\tilde{k}}(\tau)=X \cosh \tau$.
To summarize we have  the following result for $\xi_i$'s:
\begin{eqnarray} 
&
\xi_{\tilde{f}}=\half X (\cosh \tau +1), \quad
\xi_{\tilde{k}}=\half X (-\cosh \tau +1), \quad
\xi_{\tilde{F}}=- X \sinh \tau, 
&
\nonumber \\
&
\xi_{y}= 2 P X (\tau \cosh \tau - \sinh \tau), \quad
\textrm{and} \quad
\dot{\xi}_{\Phi} =0,
&
\end{eqnarray}
where the last result can be easily verified by a straightforward calculation.
Having determined the explicit form of $\xi_i$'s we can consider the equations for
the fields $\bar{\phi}^i$'s. First we write the equation for $\bar{\Phi}(\tau)$:
\begin{equation}
\dot{\Phi} = \frac{4}{3} e^{4q_0-4p_0-4Y_0} \xi_\Phi.
\end{equation}
Since $\xi_\Phi$ is constant the unique solution which is regular  at $\tau \to 0$ 
corresponds to $\xi_\Phi = 0$ and therefore $\dot{\bar{\Phi}}=0$.
For $\bar{y}$ we get:
\begin{equation}
\dot{\bar{y}} + \cosh (y_0) \bar{y} = \frac{2}{3} e^{4q_0 -4p_0-4Y_0} \xi_y.
\end{equation}
Using the result for $\xi_{y}$ and substituting the expressions for $q_0(\tau)$,
$p_0(\tau)$ and $Y_0(\tau)$ we may solve for $\bar{y}(\tau)$ :
\begin{equation}
\bar{y}(\tau) = 32\epsilon^{-8/3}  2^{2/3}  \frac{2P X}{\sinh \tau} 
                     \int_0^\tau \frac{(x \coth x -1) \sinh^2 x}{(\sinh(2x)-2x)^{2/3}} dx,
\end{equation}
where we fixed the integration constant by requiring regularity at $\tau \to 0$.
For our purposes we need an asymptotic behavior of $\bar{y}(\tau)$. At $\tau \to \infty$
we have:
\begin{equation}     \label{eq:yUV}
\bar{y} \approx \mu \left( \tau - \frac{5}{2} \right) e^{-\tau/3} + V e^{-\tau} + \ldots,
\end{equation}
where $\mu=48 \epsilon^{-8/3}  2^{1/3} 2P g_s X$ and $V$ is a
numerical constant proportional to $\mu$.
Note that $\mu$ is a dimensionless parameter (the dimensions of
$\epsilon$, $P$ and $X$ are $-\frac{3}{2}$, $-2$ and $-2$ respectively).
In the IR we find:
\begin{equation}
\bar{y}(\tau) = \frac{3^{2/3}}{27} \mu \tau^2 + O(\tau^4).
\end{equation}
The equation for $\bar{p}(\tau)$ is given by:
\begin{equation}
\dot{\bar{p}} + \frac{1}{5} \sinh (y_0) \bar{y} + 2 e^{-10 p_0} \bar{p} = 
    \frac{1}{30} e^{4q_0-4p_0-4Y_0} \xi_{p}.
\end{equation}
Using the result for $\bar{y}(\tau)$ and the fact that $\xi_{p}=0$ we may find the solution for 
$\bar{p}(\tau)$. Again we require regularity at $\tau \to 0$. 
We get:
\begin{equation}
\bar{p}(\tau) = \frac{1}{5} \frac{1}{\beta(\tau)}
             \int_{0}^{ \tau} \beta(\tau^\prime) \frac{\bar{y}(\tau^\prime)}{\sinh \tau^\prime} d \tau^\prime 
\quad \textrm{where} \quad
    \beta(\tau) \equiv e^{2 \int_{\tau_0}^{\tau} e^{-10 p_0(x)}} dx.
\end{equation}
The solution has the following asymptotic 
behavior:
\begin{equation}
\bar{p}(\tau) \approx \frac{3}{5} \mu \left( \tau - 4 \right) e^{- 4\tau/3} \quad
\textrm{at} \quad \tau \to \infty
\end{equation}
and
\begin{equation}
\bar{p}(\tau)  = \frac{3^{2/3}}{675} \mu \tau^2 + O(\tau^4) \quad
\textrm{at} \quad \tau \to 0.
\end{equation}
Next we consider the equation for $(\bar{Y}-\bar{q})$:
\begin{equation}
(\dot{\bar{Y}}-\dot{\bar{q}}) - \frac{1}{5} \sinh(y_0) \bar{y} + \frac{4}{3} e^{-10p_0} \bar{p} = 
     - e^{4q_0-4p_0-4Y_0} \left( \frac{1}{18} \xi_{Y} + \frac{1}{45} \xi_{q} \right). 
\end{equation}
In this case we fix the integration constant requiring that the function vanishes at infinity.
The result is:
\begin{equation}
\bar{Y}(\tau) - \bar{q}(\tau)  = \int_{\tau}^\infty \left( \frac{1}{5}  \frac{\bar{y}(x)}{\sinh x} +
                                        \frac{4}{3} e^{-10 p_0(x)} \bar{p}(x) \right) dx,
\end{equation}
so that:
\begin{equation}
\bar{Y}(\tau) - \bar{q}(\tau)  \approx \frac{9}{10} \mu \left( \tau - \frac{11}{4} \right) e^{- 4\tau/3} \quad
\textrm{at} \quad \tau \to \infty
\end{equation}
and
\begin{equation}
\bar{Y}-\bar{q} = C^{0}_{Y}- 3^{2/3} \frac{7}{1350} \mu \tau^2 + O(\tau^4) \quad
\textrm{at} \quad \tau \to 0,
\end{equation}
where $C^0_{Y}$ is a numerical constant proportional to $\mu$.
Now we are in a position to write down the equations for the 3-form fields. 
 Using the expressions for $\xi_{\tilde{f} \pm \tilde{k}}$
and $\xi_{\tilde{F}}$, passing from  $\tilde{f}$,  $\tilde{k}$
and $\tilde{F}$ to $f$, $k$ and $F$ we obtain and recalling that $\bar{\Phi}=0$:
\begin{eqnarray} \label{eq:fkF2}
\dot{\bar{f}\,\,} + e^{2y_0(\tau)} \bar{F} - 2 \dot{f}_0(\tau) \bar{y}  &=& - \frac{2X}{2P} h_0(\tau) (\cosh \tau -1) 
\nonumber \\
\dot{\bar{k}} - e^{-2y_0(\tau)} \bar{F} + 2 \dot{k}_0(\tau) \bar{y} &=&  \frac{2X}{2P} h_0(\tau) (\cosh \tau +1)
\nonumber \\
\dot{\bar{F}\,} - \half( \bar{k} -\bar{f} )&=& - \frac{2X}{2P} h_0(\tau) \sinh \tau.
\end{eqnarray}
Before discussing the explicit solution of this system it is worth to re-derive these equations
using the 2nd order type IIB equations of motion. In the most general ansatz preserving the global symmetry  
the 5-form $\tilde{F}_5$ is given by
\begin{equation}            
\tilde{F}_5 = \frac{1}{g_s} (1 + \star_{10}) \mathbf{d}\varphi \wedge \mathbf{d}x_0 \wedge \ldots
                               \wedge \mathbf{d}x_3,
\end{equation}
where $\varphi=\varphi(\tau)$. Supersymmetry requires
$\varphi=h^{-1}$ (see \cite{Grana:2000jj} and \cite{Gubser:2000vg}) , but it does not necessarily 
hold in a non-supersymmetric case. In what follows we will demonstrate how
assuming that $\dot{\Phi}=0$ and $\varphi=h^{-1}$ one may reproduce (\ref{eq:fkF2})
from the usual 2nd order 3-forms equations of
motion.  Indeed, under these assumptions the type IIB 3-forms equations
reduce to (\ref{eq:F3H3}). Let us expand (\ref{eq:F3H3}) around the supersymmetric KS solution. 
Note that the expansion includes also $\star_6$ due to the deformation of the 6d space.
We will denote the modified Hodge star operation by $\star_6 = \star_6^{(0)} + \bar{\star}_6$, where 
$\star_6^{(0)}$ corresponds to the supersymmetric configuration.
After some algebra the linearized RR 3-form equation reduces to:
\begin{equation}   \label{eq:Z1}
\mathbf{d}Z_3 =0,
\end{equation}
where
\begin{eqnarray}   \label{eq:Z3def}
Z_3 &=& h_0^{-1} \left( \bar{H}_3 + \star_6^{(0)}  \bar{F}_3 +  \bar{\star}_6 F^{(0)}_3 \right) =
\nonumber \\
 &=& h_0^{-1} \bigg[
     \left( \dot{\bar{f}\,\,} + e^{2y_0} \bar{F} - 2 \dot{f}_0 \bar{y}  \right)
                                                               d \tau \wedge g^1 \wedge g^2 +
       \left( \dot{\bar{k}} - e^{-2y_0} \bar{F} + 2 \dot{k}_0 \bar{y}  \right)
                                                               d \tau \wedge g^3 \wedge g^4 +
\nonumber\\ 
&&    +  \left( \half( \bar{k} -\bar{f} ) -\dot{\bar{F}\,} \right)
                        g^5 \wedge \left( g^1 \wedge g^3 + g^2 \wedge g^4 \right) \bigg].
\end{eqnarray}
where $F_3^{(0)}$ is the RR 3-form in the KS background.
In deriving this result it is convenient to use the representation (\ref{eq:6dmn})
of the 10d metric. In particular the determinant of the 6d metric is given by $g_6 = e^{2m+n}$. 
Similarly, from the NSNS 3-form equation we have:
\begin{equation}   \label{eq:Z2}
\mathbf{d} \star_6 Z_3 =0.
\end{equation}
Comparing (\ref{eq:Z3def}) with the l.h.s. of (\ref{eq:fkF2}) we might
conclude that the r.h.s. of  (\ref{eq:fkF2}) yields expressions for 
the components of the closed (and co-closed) form $Z_3$. 
Notice that having $Z_3 \neq 0$ necessary means that the complex form $G_{3}= F_{3}+ \frac{i}{g_s} H_{3}$
is not imaginary self dual and therefore the supersymmetry is broken \cite{Grana:2000jj}, \cite{Gubser:2000vg}.
The most general solution of (\ref{eq:Z1})
and (\ref{eq:Z2})
has 3 integration constants and it appears in (\ref{eq:fkFsol}). On viewing 
(\ref{eq:fkFsol}) we may conclude that the 3-form on the r.h.s. of (\ref{eq:fkF2}) 
corresponds to the divergent $(0,3)$-form we have mentioned in the discussion following (\ref{eq:fkFsol}).
 Remarkably, this is the only solution for $Z_3$, which is consistent with $\dot{\Phi}=0$.
This is evident from the linearized version of the dilaton equation of motion.
For $\dot{\Phi}=0$ it reads:
\begin{eqnarray}
F_3^{(0)} \wedge Z_3 &=& 
\frac{2X}{2P} \mathbf{M} \Big[ (-\cosh (\tau) + 1) (1-F_0(\tau)) + (\cosh (\tau) + 1) F_0(\tau) +
\nonumber \\
&&   \qquad \qquad   +  2 \sinh (\tau) F_0^\prime(\tau)    \Big] d \tau
               \wedge g^1 \wedge \ldots \wedge g^5 =0,
\end{eqnarray}
as can be verified by using an explicit expression for $F_0(\tau)$.
To find the solution for $\bar{F}(\tau)$,  $\bar{f}(\tau)$ and $\bar{k}(\tau)$
note that we already know the solution of the homogeneous part of (\ref{eq:fkF2}).
One can read these solutions from (\ref{eq:fkFsol}). Let us consider the solution
of the inhomogeneous equations in the form:
\begin{eqnarray}    \label{eq:lambdaFkf}
\bar{f}(\tau) &=& \lambda_1(\tau) f_1(\tau) +  \lambda_2(\tau) f_2(\tau) +  \lambda_3(\tau) f_3(\tau)
\nonumber \\
\bar{k}(\tau) &=& \lambda_1(\tau) k_1(\tau) +  \lambda_2(\tau) k_2(\tau)+   \lambda_3(\tau) k_3(\tau)
\nonumber \\
\bar{F}(\tau) &=& \lambda_1(\tau) F_1(\tau) +  \lambda_2(\tau) F_2(\tau) +  \lambda_3(\tau) F_3(\tau),
\end{eqnarray}
where $F_i(\tau)$, $f_i(\tau)$ and $k_i(\tau)$ (for $i=1,2,3$)   appear in (\ref{eq:fkFsol}) multiplied by $C_i$, 
for example, $F_2(\tau)=(\sinh \tau)^{-1}$ and $F_3(\tau)=0$.
Plugging this into  (\ref{eq:fkF2}) we obtain a set of linear equations for $\dot{\lambda}(\tau)_i$'s.
Solving it we get a solution for $\bar{F}(\tau)$,  $\bar{f}(\tau)$ and $\bar{k}(\tau)$.
The final expressions, which  are  quite complicated appear in
Appendix \ref{section:Tesf}.
Instead we will give the asymptotic solutions at $\tau \to \infty$ and  $\tau \to 0$. 
In the UV we have:
\begin{eqnarray}
\bar{F}(\tau) & \approx & \mu \left( \frac{3}{4} \tau - 3 \right) e^{-\tau/3} + 
                               \left( \frac{3}{2} V + V^\prime \right) e^{-\tau} + O(e^{-4\tau/3})
\nonumber \\
\bar{f}(\tau) & \approx & - \frac{27}{16} \mu e^{-\tau/3} + 
                               \left( \frac{V}{2} + V^\prime \right) e^{-\tau} + O(e^{-4\tau/3})
\nonumber \\
\bar{k}(\tau) & \approx &  \frac{27}{16} \mu e^{-\tau/3} - 
                               \left( \frac{V}{2} + V^\prime \right) e^{-\tau} + O(e^{-4\tau/3})
\nonumber \\
\bar{f}(\tau) + \bar{k}(\tau) & \approx & 
            \mu \left( -3 \tau^2 + \frac{9}{2} \tau +\frac{51}{8} \right) e^{-4\tau/3} + O(e^{-2\tau}),
\end{eqnarray}
where $V^\prime$ is a constant proportional to $\mu$.
and in the IR:
\begin{eqnarray}
\bar{k}(\tau) \approx -2 \gamma \tau + O(\tau^3),
\quad
\bar{f}(\tau) \approx  \frac{1}{2} \gamma  \tau^3+ O(\tau^5),
\quad
\bar{F}(\tau) \approx  \gamma \tau^2+ O(\tau^4),
\end{eqnarray}
where
\begin{equation}
\gamma =  - \frac{X}{3P} (4 P)^2 2^{2/3} \varepsilon^{-8/3} a_0
 =  - \frac{2^{1/3}}{18} \mu a_0.
\end{equation}
Here we used the fact that 
$h_0(\tau) \approx (4 P)^2 2^{2/3} \varepsilon^{-8/3} 
\left(a_0 - a_1 \tau^2 + \ldots \right)$.
Finally, we arrive at the last equation for the fields:
\begin{eqnarray} \label{eq:-2Y5q}
-2 \dot{\bar{Y}}+ 5 \dot{\bar{q}\,} &=& \sqrt{3} e^{-4p_0 -6q_0} \bigg( -(4 \bar{p} + 6\bar{q}) \tilde{L}_0 + 
                                         (2 P - \tilde{F}_0) \bar{\tilde{f}\,\,} +  \tilde{F}_0 \bar{\tilde{k}} +
\nonumber \\
&&                  + ( \tilde{k}_0 - \tilde{f}_0) \bar{\tilde{F}\,} \bigg)
                       + \frac{1}{9} \left( \xi_Y + \xi_q \right) e^{4q_0-4p_0 -4Y_0}.
\end{eqnarray}
We have already seen that for $\xi_Y=\xi_q=0$ the self dual 5-form $\tilde{F}_5$ is given
by (\ref{eq:5form}) like in the supersymmetric background. 
The warp function $h(\tau)$ in this case is given by (\ref{eq:dh}).  Linearizing   (\ref{eq:dh})
around the supersymmetric configuration and recalling that $h \sim e^{10q-4Y}$
we may re-derive (\ref{eq:-2Y5q})  for the special case $\xi_Y=\xi_q=0$ . This equation
is easily solved once we use the expression for $\bar{Y}-\bar{q}$
(see Appendix \ref{section:Tesf} for the full solution). We obtain:
\begin{equation}
\bar{q}(\tau)  \approx-\frac{2}{5} \mu  \tau  e^{- 4\tau/3} + O\left(e^{- 4\tau/3}\right)\quad
\textrm{at} \quad \tau \to \infty
\end{equation}
and
\begin{equation}
\bar{q} =  3^{2/3} \frac{11}{4050} \mu \tau^2 + O(\tau^4) \quad
\textrm{at} \quad \tau \to 0,
\end{equation}
This completes our solution for various fields in the non-supersymmetric background. 
The deformation is controlled by the single parameter $\mu$ and all the fields have a regular behavior
in the UV and in the IR. There are two non-normalizable modes.
The first one is $y(\tau)$ and it is related to the deformation of the 6d metric. The second one 
is $\xi_2$ and it is associated with the 3-forms. In the UV we have:
\begin{equation}
\xi_2 \sim -F+\frac{k-f}{2} \approx -\frac{3}{4} \mu \left( \tau - \frac{25}{4} \right) e^{-\tau/3}.  
\end{equation}
Both $y(\tau)$ and  $\xi_2$ have dimension $\Delta=3$ which matches perfectly with the
asymptotic behavior of the fields. In the dual gauge theory these operators are dual to the gaugino 
bilinears. The deformation also involves other fields like $s=f+k$ with a normalizable 
behavior at $\tau \to \infty$. For example, $s \approx e^{-4 \tau/3}$ as expected for
an operator with $\Delta=4$.

Notice also that the field $\bar{Y}$ does not vanish at $\tau=0$
(namely, $\bar{Y}=C_Y + O(\tau^2)$).
This is in contrast to the IR solution of \cite{Apreda:2003gc}.
We will return to this point in Section \ref{section:Tpwl}.

\section{Vacuum energy}
\label{section:Ve}

To calculate the vacuum energy of the deformed non-supersymmetric theory 
we will use the standard AdS/CFT technique \cite{Aharony:1999ti}.
The supergravity dual of the gauge theory Hamiltonian is a $G_{00}$
component of the 10d metric. The vacuum energy, therefore, can be found
by variation of the type IIB SUGRA action (see Appendix \ref{section:TIIBeom}) with respect to $G_{00}$.
This variation vanishes on-shell, except a boundary term. Looking at the supergravity
action, it is clear that the only such a boundary term will appear from the curvature part
of the action. Since the vacuum energy does not depend on the world-volume coordinates
we might consider the metric variation in the form:
\begin{equation}
G_{00} \to q G_{00}.
\end{equation}   
Under this variation  the Christoffel connection symbols transform as:
\begin{equation}
\delta \Gamma_{00}^\tau = -\half q G^{\tau\tau} \partial_\tau G_{00}
\quad \textrm{and} \quad
\delta \Gamma_{0\tau}^0 =  q G^{00} \partial_\tau G_{00}
\end{equation}
and \footnote{here we use the formula:
          \begin{equation}
            \delta \left( \sqrt{-G}\mathcal{R} \right) 
                     =G^{\mu\kappa}G^{\lambda\nu} \delta
                     G_{\kappa\lambda} R_{\mu\nu} \sqrt{-G} + \half
                     \sqrt{-G}\mathcal{R} G^{\mu\nu} \delta G_{\mu\nu}+
                       \partial_\kappa \left(  G^{\mu\nu} \delta
                          \Gamma^\kappa_{\mu\nu}\sqrt{-G}  \right) -
                       \partial_\nu \left(  G^{\mu\nu} \delta
                          \Gamma^\kappa_{\kappa\mu}\sqrt{-G}  \right)                       
                           .
          \end{equation}    }:
\begin{equation}
\delta \left( \sqrt{-G}\mathcal{R} \right) = \ldots - 
  q \partial_\tau \left[ \half \sqrt{-G} G^{\tau\tau} G^{00} \partial_\tau G_{00} \right],
\end{equation}
where $(\ldots)$ denotes other non-boundary terms which are canceled on-shell by 
terms coming from the forms part of the action.
Note that unlike \cite{Evans:2002mc} in our case there is no additional boundary
term since the dilaton is taken to be constant. 
Substituting the on-shell values of the 10d metric  
($G_{00}=-h^{-1/2}(\tau)$,
$G_{\tau\tau} \sim h^{1/2}(\tau) e^{m(\tau) -\half n(\tau)}$ and
$\sqrt{-G} \sim h^{1/2}(\tau) e^{m(\tau) +\half n(\tau)} $) we obtain an expression for the vacuum energy:
\begin{equation}
E \sim \lim_{\tau \to \infty} \left( e^{n(\tau)} \partial_\tau \ln h(\tau) \right).
\end{equation}
The divergent result we have found is expected to 
be canceled out  when we compare the vacuum energies of
our solution and of the KS background, which we take as a reference. Using that
$h \to h_0 + \bar{h}$  and $n \to n_0 + \bar{n}$  
we get:
\begin{equation}
\Delta E \sim \left[ e^{n_0} \left(   \partial_\tau \left(\frac{\bar{h}}{h_0}\right) 
                                    +\bar{n}  \partial_\tau \left(\ln h_0\right)  \right) 
                \right]_{\tau \to \infty} 
\end{equation} 
The connection between $\bar{n}$ and $\bar{h}$ and the fields we have found in the previous section
is:
\begin{equation}
\bar{n} = -4 \bar{q} + 4 \bar{p} + 4 \bar{Y}
\quad \textrm{and} \quad
\frac{\bar{h}}{h_0} = 10 \bar{q} - 4 \bar{Y}.
\end{equation}
Therefore
\begin{equation}  \label{eq:Dexxx}
\Delta E \sim \left[ e^{n_0} \left(  \left( 10 \dot{\bar{q}} - 4 \dot{\bar{Y}} \right)
                                       -\frac{4}{3} \left(  -4 \bar{q} + 4 \bar{p} + 4 \bar{Y} \right) \right)
                                    \right]_{\tau \to \infty}
  \sim \mu, 
\end{equation} 
where we used the asymptotic solutions for the fields at $\tau \to
\infty$ from the previous section.
In (\ref{eq:Dexxx}) the term $e^{n_0(\tau)}$ diverges at $\tau \to \infty$ as $e^{4\tau/3}$. This is
suppressed by the $e^{-4\tau/3}$ term in the large $\tau$ expansion of
the fields appearing in the parenthesis  which multiply  the
$e^{n_0(\tau)}$ term. 
Furthermore, the term linear at $\tau$ cancels and we end up with a
constant proportional to $\mu$.

\section{Dual gauge theory}
\label{section:Dgt}

As was announced in the introduction the deformation of the supergravity
background corresponds in the gauge theory to 
an  insertion of the soft supersymmetry breaking gaugino mass terms.
The most general gaugino bilinear term has the form of
$\mu_+ {\cal O}_+ +\mu_-{\cal O}_- + c.c$ where ${\cal O}_\pm\sim
Tr[W_{(1)}^2 \pm W_{(2)}^2]$ and $W_{(i)}, \ i=1,2 $ relate  to the
$SU(N+M)$ and $SU(N)$ gauge groups respectively. Namely, the general
deformation is characterized by two complex masses. Our
non-supersymmetric deformation of the KS solution derived above
is a special case that depends on only one real parameter $\mu$.
Since the supergravity identification of the operators ${\cal O}_\pm$
is known up to some constants of proportionality we can not determine
the precise form of the soft symmetry breaking term. 

In the non-deformed supersymmetric 
theory the $U(1)_R$ symmetry is broken \cite{Klebanov:2002gr}, \cite{Herzog:2002ih}
first by instantons to $\mathbf{Z}_{2M}$ and then further
spontaneously broken down to
$\mathbf{Z}_2$
by a VEV of the gaugino bilinear. Let us discuss first the latter
breaking. 
We have already seen that on the SUGRA side this fact is manifest from 
the UV behavior of the complex 3-form $G_{3}= F_{3}+ \frac{i}{g_s} H_{3}$.
The sub-leading term in the expansion of $G_{3}$ preserves only the $\mathbf{Z}_2$
part of the $U(1)_R$ symmetry and it vanishes at infinity like $e^{-\tau}$ matching the 
expectation from the scalar operator $\textrm{Tr}(\lambda\lambda)$ of
dimension 3 with a non-zero VEV
\cite{Loewy:2001pq}.
Plugging the non-supersymmetric solution into $G_{3}$ we find that the leading term breaking
the  $U(1)_R$ symmetry behaves like $\Delta G _3 = g(\tau) e^{-\tau/3}$, where $g(\tau)$
is some polynomial in $\tau$. This is exactly what one would predict for an operator with
$\Delta=3$ and a non-trivial mass. The second combination of the gaugino bilinears
is encoded in the 6d part of the metric. For the 6d metric in (\ref{eq:6dmn}) to preserve
the $U(1)_R$ one has to set $y=0$. In the supersymmetric deformed conifold metric 
$y(\tau)= -2 e^{-\tau} + \ldots$ similarly to the behavior of the 3-form.
In the non-supersymmetric solution $y(\tau)$ goes like $e^{-\tau/3}$ elucidating again
that the gaugino combination gets a mass term. Notice also that the non-zero
VEVs of the gaugino bilinears are modified by the SUSY
breaking deformation. This is evident, for example, from the
$Ve^{-\tau}$ term in  the UV
expansion of $\bar{y}(\tau)$ in (\ref{eq:yUV}). Clearly, for $V \neq 0$
we have a correction to the VEV in the supersymmetric theory which was
encoded in the expansion of $y_0(\tau)$. Similar $e^{-\tau}$ term appears also in the
expansion of $\xi_2(\tau)$ and therefore the VEV of the second combination
of the gauginos gets modified too.

The  spontaneous  breaking of the $\mathbf{Z}_{2M}$ discrete group   down to the $\mathbf{Z}_2$ subgroup by
gaugino condensation results in an  
 $\mathbf{M}$-fold degenerate   vacua.
 This degeneracy is generally lifted by soft breaking mass terms in the action.
For small enough masses one can treat the supersymmetry breaking as a perturbation
yielding (for a single gauge group)  the well-known result \cite{Masiero:1985ss}
that the difference in energy between a non-supersymmetric
solution and its supersymmetric reference is given by:
\begin{equation}
\Delta E \sim \textrm{Re} (\mu C),
\end{equation}
where $\mu$ and $C$ are the mass and the gaugino condensate respectively. 
For the gauge theory dual of the  deformed KS solution the vacuum
energy will in general be proportional to $\textrm{Re} (a_+\mu_+ C_+
+a_-\mu_-C_- )$
where $C_\pm$ are  the expectation values of ${\cal O}_\pm$ and
$a_\pm$ are some proportionality constants. In the special deformation
we are discussing in this paper this reduces to $\mu \textrm{Re} (a_+ C_+
+a_-C_- )$.
 In the previous section we have derived a result using the SUGRA dual
 of the gauge theory
which has this structure.
For the softly broken MN background similar calculations were
performed by \cite{Evans:2002mc}. In their case the explicit linear
dependence on the condensate was demonstrated.

One of the properties of the supersymmetric gauge theory is 
the space-time independence of the correlation function of two gaugino bilinears.
This appears from the supergravity dual description  as follows \cite{Loewy:2001pq}.
Consider a perturbation of the complex 2-form $C_2 = C^{RR}_2 + i B^{NS}_2$:
\begin{equation}
C_2 \to C_2 + y(x,\tau) \omega_2
\quad \textrm{and}
\quad
G_3 \to G_3 + y(x,\tau) \wedge \omega_3 + \mathbf{d} y(x,\tau) \wedge \omega_2,
\end{equation}
where $\omega_{2,3}$ are given by (\ref{eq:DC}) and  (\ref{eq:DG})
and  $y(x,\tau)$ has non-vanishing boundary values.
Plugging this forms into the relevant part of the type IIB action:
\begin{equation}
\int dx d \tau \sqrt{g} \left[ G_3 G_3^\star + \left( F_5 - \frac{1}{2 i} 
(C_2 \wedge G_3^\star - C_2^\star \wedge G_3 ) \right) \right]
\end{equation}
and integration over $\tau$ will \emph{not} lead to a kinematic term $dy(x_1)dy(x_2)$
and therefore the corresponding correlation function will be space-time independent.
This derivation is only schematic since there is a mixing between the 3-form modes and
the modes coming from metric as we have seen in Section \ref{section:NseoKS}.
Notice, however that this simplified calculation will yield the kinetic term for
the deformed non-supersymmetric background, since 
the complex 3-form is not imaginary self dual in this case. Thus in the non-supersymmetric 
theory the correlation function will be time-space dependent as one would expect.

\section{The plane wave limit} 
\label{section:Tpwl}

In this section we will construct a Penrose limit of the non-supersymmetric background. 
Following \cite{Gimon:2002nr} we will expand the metric around a null geodesic that
goes along an equator of the $S^3$  at $\tau = 0$.
The parameter $\varepsilon$ appearing in the 6d metric of the deformed conifold
and the gauge group parameter $\mathbf{M}$ are both taken to infinity in the PL limit,
while keeping finite the mass of the glue-ball:
\begin{equation}
M_{gb} \sim \frac{\varepsilon^{2/3}}{g_s \mathbf{M} \alpha^\prime}.
\end{equation}
Let us start with the following general ansatz of a 10d metric:
\begin{eqnarray}
ds^2 &=& D^{-1}(\tau) \left( dt^2 + dx_i^2 \right) + 
\nonumber \\
&&   + D(\tau) \left( A(\tau) \left(d \tau^2 + g_5^2 \right) + B(\tau) \left(g_3^2 + g_4^2 \right) +
     C(\tau) \left(g_1^2 + g_2^2 \right)  \right),
\end{eqnarray}
where
\begin{eqnarray}
&
A(\tau)= A_0 + A_1 \tau^2 + \ldots   \quad
B(\tau)= 2A_0 + B_1 \tau^2 + \ldots  \quad
C(\tau)= \half A_0 \tau^2 + \ldots   \quad
&
\nonumber \\
& D(\tau)= D_0 + D_1 \tau^2 + \ldots &.
\end{eqnarray}
It can be easily verified that the 10d metric in the KS solution and its non-supersymmetric deformation
have this form near $\tau =0$.
Since we expand the metric around the equator of the $S^3$
it will be useful to switch to a basis of one forms $\omega_1$,  $\omega_2$,  $\omega_3$ 
and two additional angles $\theta$ and $\phi$, related to the 1 forms
$g_i$'s by \cite{Gimon:2002nr}:
\begin{eqnarray}
g_5 &=& \sin \theta \cos \phi \omega_1 - \sin \theta \sin \phi \omega_2 + \cos \theta \omega_3
\nonumber \\
\cos(\psi/2) g_1 + \sin(\psi/2) g_2 & = & \frac{1}{\sqrt{2}} \left(
           \cos \theta \cos \phi \omega_1 - \cos \theta \sin \phi \omega_2 - \sin \theta \omega_3
           - 2 \sin \theta d \phi \right)
\nonumber \\
-\sin(\psi/2) g_1 + \cos(\psi/2) g_2 & = & - \frac{1}{\sqrt{2}} \left(
            \sin \phi \omega_1 + \cos \phi \omega_2 - 2 d \theta \right)
\nonumber \\
\cos(\psi/2) g_3 + \sin(\psi/2) g_4 & = & \frac{1}{\sqrt{2}} \left(
           \cos \theta \cos \phi \omega_1 - \cos \theta \sin \phi \omega_2 - \sin \theta \omega_3 \right)
\nonumber \\
-\sin(\psi/2) g_3 + \cos(\psi/2) g_4 & = & - \frac{1}{\sqrt{2}} \left(
            \sin \phi \omega_1 + \cos \phi \omega_2 \right).
\end{eqnarray}
In terms of the $S^3$ angle coordinates $(\theta^\prime,\phi^\prime,\psi^\prime)$ 
the geodesic lies at $\theta^\prime=0$
and is generated by $\phi_+= \half (\phi^\prime +\psi^\prime)$. Under re-scaling $\theta^\prime \to \theta^\prime/L $
the 1-forms $\omega_1$ and  $\omega_2$ will go like $1/L$ and for $\omega_3$ we obtain:
\begin{equation}
\omega_3 = 2 d \phi_+ -\half \left( \frac{\theta^\prime}{L} \right)^2 d \phi^\prime. 
\end{equation} 
In order to take the Penrose limit we define new coordinates:
\begin{eqnarray}
u = \left( D_0 A_0\right)^{1/2} \tau \sin \theta e^{i (\phi + \phi_+)} \qquad
z = \left( D_0 A_0\right)^{1/2} \tau \cos \theta  
\nonumber \\
v = \left( D_0 A_0\right)^{1/2} \theta^\prime  e^{i (\phi^\prime - \phi_+)}  \qquad
x_i \to \frac{x_i}{D_0^{1/2}}
\nonumber \\
\end{eqnarray}
together with
\begin{equation}
t = x_+ 
\quad \textrm{and} \quad
\phi_+ = \frac{1}{2 D_0 A_0^{1/2}} \left( x_+ - 2 D_0 x_- \right).
\end{equation}
Finally, re-scaling $D_0 \to D_0 L^2$ and $A_0 \to A_0 L^{-4}$ and taking $L \to \infty$
we arrive at the following pp-wave metric:
\begin{eqnarray}
ds^2 &=& - 4 dx_- dx_+ - m_0^2 \left( v \bar{v} +
                                  \left( -\frac{4A_1}{A_0} -\frac{8D_1}{D_0}  \right) z^2 +
                                  \left( -\frac{2B_1}{A_0} -\frac{8D_1}{D_0}  \right) u \bar{u}\right) dx_+^2 +
\nonumber \\
&&           + dx_i^2 + dz^2 + du d\bar{u} + dv d\bar{v},
\end{eqnarray}
where $m_0=(2 D_0 A_0^{1/2})^{-1}$ remains finite in the $L \to \infty$ limit. Now we are in a position
to take the PL of the non-supersymmetric background we have found in Section 3. Note that:

\begin{equation}
D(\tau) = h^{1/2}(\tau), \quad A(\tau)=e^{m(\tau)-\half n(\tau)},\quad
B(\tau)=e^{\half n(\tau)-y(\tau)},\quad
C(\tau)=e^{\half n(\tau)+y(\tau)},\quad
\end{equation}
and the relations between $h(\tau)$,$m(\tau)$, $n(\tau)$
 and the fields $Y(\tau)$, $p(\tau)$ and $q(\tau)$
are given in (\ref{eq:mnhYpq}).  The final result is:
\begin{eqnarray}
ds^2 &=& - 4 dx_- dx_+  + dx_i^2 + dz^2 + du d\bar{u} + dv d\bar{v} +
\nonumber \\
   &&    - m_0^2 \Bigg[ v \bar{v} +
                                  \left( \left( \frac{4a_1}{a_0} -\frac{4}{5}\right) 
                                                 - 8\frac{3^{2/3}}{135} \mu  \right) z^2 +
\nonumber\\
   &&       \qquad                    +  \left( \left( \frac{4a_1}{a_0} -\frac{3}{5}\right)
                                                 + 4\frac{3^{2/3}}{135} \mu \right) u \bar{u} \Bigg] dx_+^2,
\end{eqnarray}
where
\begin{equation}          \label{eq:m0}
m_0^2 = \frac{3^{1/3} \varepsilon^{4/3}}{2 (g_s \mathbf{M} \alpha^\prime)^2 a_0} \left(1 + 2C_Y \right).
\end{equation}
Recall that $C_Y$ is a numerical constant proportional to $\mu$. As expected for $\mu=0$ we recover the result of
the supersymmetric case \cite{Gimon:2002nr}.  
We see that all the world-sheet masses ($m_v$,$m_z$ and $m_u$) depend on
the supersymmetry breaking parameter.
Under the Penrose limit the 3-forms read:
\begin{eqnarray}
F_3 &=& \frac{3 i m_0}{\sqrt{2} g_s} \left(\frac{a_1}{a_0}\right)^{1/2} 
     dx_+ \wedge \left( \left( \frac{1}{3} + 4 \gamma) du \wedge d \bar{u} + dv \wedge d \bar{v} \right) \right) 
\nonumber \\
H_3 &=& \frac{i m_0}{\sqrt{2}} \left(\frac{a_1}{a_0}\right)^{1/2} 
     dx_+ \wedge (1-6 \gamma) \left(  du \wedge d \bar{v} - d \bar{u} \wedge d v \right)
\end{eqnarray}
and the complex 3-form is given by:
\begin{eqnarray}
G_3 = F_3 + \frac{i}{g_s} H_3 &=&
 \frac{i m_0}{\sqrt{2} g_s} \left(\frac{a_1}{a_0}\right)^{1/2} 
 dx_+ \wedge \Bigg[\left( (1 + 12 \gamma)  du \wedge d \bar{u} + dv \wedge d \bar{v} \right) +
\nonumber \\
&&  \qquad
   +   i (1 - 6 \gamma)  \left(  du \wedge d \bar{v} - d \bar{u} \wedge d v \right)
                 \Bigg],
\end{eqnarray}
where $\gamma = - \frac{1}{18} 2^{1/3} \mu a_0$.

As a non-trivial check of our solution we can verify that the equation of motion:
\begin{equation}
R_{++} = \frac{g_s^2}{4} \left( G_3 \right)_{+ij}  \left( \bar{G}_3 \right)_+^{\,\,ij} 
\end{equation}
is satisfied. Indeed:
\begin{equation}
R_{++} = m_0^2 \left[ 2 +\left( \left( \frac{4a_1}{a_0} -\frac{4}{5}\right) - 8\frac{3^{2/3}}{135} \mu \right) +
                     2 \left( \left( \frac{4a_1}{a_0} -\frac{3}{5}\right) + 4\frac{3^{2/3}}{135} \mu \right)
               \right] = 12  m_0^2 \, \frac{a_1}{a_0}
\end{equation}
and the norm of the 3-form is:
\begin{equation}
 \left( G_3 \right)_{+ij}  \left( \bar{G}_3 \right)_+^{\,\,ij} = m_0^2 \, \frac{a_1}{a_0} \,
  4 \left( (1+12 \gamma)^2 + 9 + 2 (1-6 \gamma)^2 \right) = 48 m_0^2 \, \frac{a_1}{a_0} + O(\gamma^2)
\end{equation}
matching perfectly with the $R_{++}$. Notice that in the last equation we have neglected 
corrections of the 2nd order in the deformation.

\section{The plane wave string theory and  the Annulons} 
\label{section:TplwtatA}

The  string  theory  associated with the plane wave background
described in the previous section 
is quite  similar to that associated with the PL limit of the 
KS background.  The bosonic sector includes three massless fields that correspond to the 
spatial directions on the world-volume of the D3 branes. Their masslessness is 
attributed to the translational invariance of the original metric and
the fact that
the null geodesic is at constant $\tau$. The rest five coordinates
are massive. 
Altogether the bosonic spectrum takes the form

\begin{eqnarray}
& \omega_n^i = n \quad \textrm{for} \quad i=1,2,3;  \qquad
  \omega_n^z = \sqrt{n^2 + \hat{m}_z^2};   &
\nonumber \\
& \omega_n^{u,v} = \sqrt{n^2 + \half \left( \hat{m}_v^2 + \hat{m}_u^2 \right) \pm 
             \sqrt{ \frac{1}{4} \left( \hat{m}_v^2 - \hat{m}_u^2 \right)^2 + n^2 \hat{m}_B^2 }};  &
\end{eqnarray}
where 
\begin{eqnarray}
&
\hat{m}_v = p^+ \alpha^\prime m_0, \quad
\hat{m}_B =   \sqrt{2} p^+ \alpha^\prime m_0 \left( \frac{a_1}{a_0} \right)^{1/2} (1-6\gamma), 
&
\nonumber \\
&
\hat{m}_z = p^+ \alpha^\prime m_0 \sqrt{ \frac{4a_1}{a_0} -\frac{4}{5} - 8\frac{3^{2/3}}{135} \mu}
\quad \textrm{and} \quad
\hat{m}_u = p^+ \alpha^\prime m_0 \sqrt{ \frac{4a_1}{a_0} -\frac{3}{5} + 4\frac{3^{2/3}}{135} \mu}.
&
\end{eqnarray}
The difference between the bosonic spectrum of the 
deformed model and that of \cite{Gimon:2002nr} is the shift of the masses
of the $z,v,\bar v,u,\bar u$ fields. 
The sum of the $mass^2$ of the individual fields $\sum m^2 = 12m_0^2  \frac{a_1}{a_0}$ has  the same 
form as the sum in the supersymmetric case apart from the modification of $m_0$ (\ref {eq:m0}).
The modification of $m_0$  is also responsible for   the deviation of
the deformed string tension with from  the supersymmetric one since
the string tension 
$T_s \sim g_s \mathbf{M}   m_0^2$.

The fermionic spectrum takes the form
  \begin{eqnarray}
& \omega_n^k = \sqrt{n^2 + \hat{m}_B^2 \left( \frac{1+3 \gamma}{1 - 6 \gamma}  \right)^2}
         \approx \sqrt{n^2 + \hat{m}_B^2 \left( 1 + 18 \gamma  \right)}
\quad \textrm{for} \quad k=1, \ldots, 4; &
\nonumber \\
& \omega_n^l =  
             \sqrt{n^2 + \frac{1}{4} \hat{m}_B^2} \pm \frac{1}{2} \hat{m}_B  
\quad \textrm{for} \quad l=1,2. &
\end{eqnarray}
Comparing the bosonic and fermionic masses we observe that like in  the undeformed KS model 
there is no linearly realized world-sheet supersymmetry and the hence
there is a 
non-vanishing  zero point energy.
However,  up to deviations linear in $\mu$ the sum
of the square of the frequencies  of the bosonic and fermionic modes match. 
Since this property follows  in fact from the relation between 
$R_{++}$ and $\left( G_3 \right)_{+ij}  \left( \bar{G}_3 \right)_+^{\,\,ij}$  
it should be  a universal property  of any plane wave background.  

Surprisingly we find that the fermionic spectrum admits  two fermionic zero modes $\omega_0^{l=1,2}$
exactly like in the supersymmetric case.
The fermionic zero modes in the spectrum of the latter case  
were predicted \cite{Gimon:2002nr} upon observing  that 
the Hamiltonian still commutes with the four supercharges that correspond to the four dimensional
${\cal N}=1$ supersymmetric gauge theory. 
This implies that four supersymmetries out of the sixteen
supersymmetries of plane wave solution
commute with the Hamiltonian giving rise to the four zero-frequency modes
and a four dimensional Hilbert sub-space of (two bosonic and two fermionic) degenerate states.
One might have expected that in the PL of the deformed  theory
the fermionic zero modes will be lifted by an amount proportional to the supersymmetry breaking parameter.
Our results, however, contradict this expectation.
In the dual field theory this implies that even though the full theory is non-supersymmetric,
the sector of states with large $J$ charge  admits supersymmetry. As will be
discussed below  these states
are characterized by their large mass which is proportional to $J$. Presumably, in this limit of states of large mass the 
impact of the small  gaugino mass deformations is washed away. For instance one can estimate that the ratio of the  boson 
fermion mass difference to the mass of the annulon scales like $
\frac{\mu}{J}$ 
and since $\mu$ has to be small and $J\rightarrow \infty$ this ratio is negligible.

Note that the fermionic zero modes are in accordance with the criteria   presented in
\cite{Apreda:2003gs}. However, the metric and the 3-form given
in \cite{Apreda:2003gs} do not coincide   with our results, because of the factor of $C_Y$ in the
expression for $m_0^2$.

Since apart from the modifications  of the fermionic and bosonic frequencies the string Hamiltonian we find
has the same structure as the one found for the KS case,  the analysis of the corresponding
gauge theory states also follows that of  \cite{Gimon:2002nr}. 
We will not repeat here this analysis, but rather just summarize its outcome:
\begin{itemize}
\item
The ground state of the string corresponds to the {\it Annulon}. This hadron which carries a large $J$ charge 
is also very massive since  its mass is given by 
\begin{eqnarray}
M_{annulon} = m_0 J 
 \end{eqnarray} 
Obviously, the only difference between the annulon of the deformed theory in comparison with the 
supersymmetric one is the modification of $m_0$.
\item
The annulon can be described as  a ring 
composed of $J$ constituents each having a mass ( in the mean field of all the others)  of $m_0$.
\item
The annulon which is a space-time scalar has a fermionic superpartner of the same mass. The same holds for the rest
of the bosonic states.
\item
The string Hamiltonian has a term $ \frac{P_i^2}{2m_0J}$ that describes a non-relativistic motion of the annulons.
\item
The annulons admit stringy ripples. The spacing between these
excitations are proportional 
to $\frac{T_s} {M_{annulon}}$.
\item 
The string Hamiltonian describes also excitations that correspond to
the addition of small number
 of different constituents on top of the J basic ones.
\end{itemize}

\section{Discussion}

In this paper we have found an explicit solution for the first order deformation of 
the KS supergravity background. This deformation breaks the supersymmetry as one can see,
for example, from the structure of the deformed complex 3-form, which
is not imaginary  self dual
as required by the type IIB BPS equations. We have verified that the solution is regular in the IR
and the UV and the leading order  UV behavior of various fields
matches  their conformal dimensions.
We have also identified two fields with a non-normalizable modes.
In the dual gauge theory these fields correspond to the gaugino bilinears and therefore
the deformation is related to the insertion of the softly
supersymmetry breaking gaugino mass terms.
Theses masses remove the degeneracy of the vacuum.  Using the dual supergravity description 
we have checked that the lifting of the vacuum energy satisfies the  prediction for  
$\mathcal{N}=1$ theories.
Finally, we have investigated the plane-wave limit of the non-supersymmetric background finding that 
there are two fermionic zero frequencies exactly like in the PL of the supersymmetric solution.

There are plenty of open questions that deserve further
investigation. Let us mention only few of them.
The solution we have found is by no means the most general one. It is
characterized by one real gaugino mass whereas in general one can
introduce two complex masses. It will be interesting to determine the
corresponding general solutions of the equations of motion.
In  the  laboratory of the non-supersymmetric solution we have found, it will be interesting
to ``measure'' certain properties of the gauge dynamics like Wilson
loops, 't Hooft loops, baryonic configurations, fundamental quarks via
D brane probes etc. 
Another interesting question to explore is whether, the 
surprising supersymmetry of the gauge sector dual of the annulons in
the overall non-supersymmetric theory, will survive in the presence of
$\frac {1}{J} $corrections.

\acknowledgments

We would like to thank
Yaron Oz, Leo Pando Zayas and Tadakatsu Sakai for fruitful discussions.
We would  specially like to  thank Ofer Aharony for many useful
conversations about the project and for
his illuminating  comments about the manuscript. 
This work was supported in part by the German-Israeli Foundation for
Scientific Research and by the Israel Science Foundation.

\appendix

\section{Type IIB equations of motion}
\label{section:TIIBeom}

In Einstein frame the bosonic part of the type IIB action 
is:
\begin{eqnarray}
&& \frac{1}{2 \kappa^2} \int d^{10} x \sqrt{-g} \mathcal{R}-
\frac{1}{4 \kappa^2} \int d^{10} x \Big[ 
               \mathbf{d} \Phi \wedge  \star \mathbf{d} \Phi +
               e^{2\Phi} \mathbf{d} C_0 \wedge  \star \mathbf{d} C_0 +
\nonumber \\
&& \quad
               g_s e^{-\Phi} H_3 \wedge \star H_3 
                  + g_s e^{\Phi} \tilde{F}_3 \wedge \star \tilde{F}_3 
               +  \frac{g_s^2}{2} \tilde{F}_5 \wedge \star \tilde{F}_5 
               + g_s^2 C_4 \wedge H_3 \wedge F_3 \Big].
\end{eqnarray}

The field equations are
\begin{eqnarray}
 \mathbf{d} \star \mathbf{d}  \Phi  &=&  e^{2\Phi} \mathbf{d} C_0 \wedge \star \mathbf{d} C_0 
      - \half g_s e^{-\Phi} H_3 \wedge \star H_3 + 
        \half g_s e^{\Phi} \tilde{F}_3 \wedge \star \tilde{F}_3,
                                                                    \nonumber \\
\mathbf{d} \left(  e^{2\Phi} \star \mathbf{d} C_0  \right) &=& 
                            -  g_s e^{\Phi} H_3 \wedge \star \tilde{F}_3,   \nonumber\\
\mathbf{d} \star  \left(  e^{\Phi} \tilde{F}_3 \right) &=&  g_s  F_5 \wedge  H_3,   \nonumber\\
\mathbf{d} \star  \left(  e^{-\Phi} H_3 - C_0  e^{\Phi}\tilde{F}_3 \right) &=&
                                                   - g_s  F_5 \wedge F_3,   \nonumber\\
\mathbf{d} \star \tilde{F}_5   &=&   - F_3 \wedge  H_3,
                                                                            \nonumber\\
R_{mn} &=& \half \partial_m \Phi \partial_n \Phi
             + \half  e^{2\Phi}\partial_m C_0 \partial_n C_0 
             + \frac{g_s^2}{96} \tilde{F}_{mpqrs} \tilde{F}_{m}^{\,\,\,pqrs}
                                                                            \nonumber\\
&&             + \frac{g_s}{4}\left(
                   e^{-\Phi} H_{mpq} H_{m}^{\,\,\,pq} + 
                   e^{\Phi}  \tilde{F}_{mpq} \tilde{F}_{m}^{\,\,\,pq}  
                              \right)
                                                                            \nonumber\\
&&             -  \frac{g_s}{48} g_{mn} \left(
                   e^{-\Phi} H_{mpq} H^{mpq} + 
                   e^{\Phi}  \tilde{F}_{mpq} \tilde{F}^{mpq}
                            \right). 
\end{eqnarray}
Here
\begin{eqnarray}
\tilde{F}_3 &=& F_3 - C_0 H_3, \qquad F_3 = \mathbf{d} C_2,    \nonumber\\
\tilde{F}_5 &=& F_5 - C_2 \wedge H_3, \qquad F_5 = \mathbf{d} C_4 \nonumber\\
H_3 &=& \mathbf{d} B_2.
\end{eqnarray}
The Bianchi identities are:
\begin{eqnarray}
\mathbf{d} \tilde{F}_3 &=& -\mathbf{d} C_0\wedge H_3,     \nonumber\\
\mathbf{d} \tilde{F}_5 &=& -F_3 \wedge H_3
\end{eqnarray}
and the 5-form is self dual:

\begin{eqnarray}
\star \tilde{F}_5 &=&  \tilde{F}_5.
\end{eqnarray}

\section{The explicit solutions for $f(\tau)$, $k(\tau)$, $F(\tau)$
  and $q(\tau)$}
\label{section:Tesf}

The functions $f(\tau)$, $k(\tau)$, $F(\tau)$ are given by
(\ref{eq:lambdaFkf}), where:

\begin{eqnarray}
\lambda_1 &=& \half \int_\tau^\infty \frac{k_0^\prime (x) +f_0^\prime(x) }{\sinh x} \bar{y}(x) dx
\nonumber \\
\lambda_2 &=& \half \int_0^\tau \left(
 \half \left( \cosh x -\frac{x}{\sinh x} \right) (k_0^\prime (x) +f_0^\prime(x)) \bar{y}(x) -
                    \frac{2X}{2P} h_0(x) \sinh^2 x
                                \right) dx
\nonumber \\
\lambda_3 &=& - \int_\tau^\infty \Bigg(
                   (f_0^\prime (x) - k_0^\prime(x)) \bar{y}(x) - \half (k_1(x) +f_1(x)) \lambda_1^\prime(x)
                     - \half (k_2(x) +f_2(x)) \lambda_2^\prime(x) + 
\nonumber \\
&& \qquad 
                  + \frac{2X}{2P} h_0(x)
                                \Bigg) dx  
\end{eqnarray}
For $q(\tau)$ we have:
\begin{eqnarray}
\bar{q(\tau)} &=& \frac{1}{\gamma(\tau)} \int_{0}^{\tau} \gamma(x)
  \Bigg(
          \frac{2}{3}(\dot{\bar{Y}}-\dot{\bar{q}}) 
         + \frac{\sqrt{3}}{3} e^{-4p_0(x)-6q_0(x)} \tilde{L}_0(x) 
                 \bigg(  -4 \bar{p}(x) +
\nonumber \\
&& \qquad
                                       + \frac{1-F_0(x)}{L_0(x)} \bar{f}(x) 
                                       + \frac{F_0(x)}{L_0(x)} \bar{k}(x) 
                                       + \frac{k_0(x)-f_0(x)}{L_0(x)} \bar{F}(x)  \bigg)
  \Bigg)
d x
\end{eqnarray}
with 
\begin{equation}
\gamma(\tau) = e^{2 \sqrt{3} \int_{\tau_0}^{\tau} e^{-4p_0(x)-6q_0(x)} \tilde{L}_0(x) dx} .
\end{equation}

\section{The complex $(0,3)$ form on the deformed conifold}  \label{section:03}

In this section we demonstrate by an explicit calculation that the complex
3-form $G_3 \equiv F_3 + i H_3$ given by (\ref{eq:F3}), (\ref{eq:H3})
and (\ref{eq:fkFsol}) for $C_1=-\half$ and $C_2=C_3=0$ is a $(0,3)$
form on the deformed conifold space. In deriving this result
we will use the complex structure of the deformed conifold
identified in \cite{Papadopoulos:2000gj} (one might alternatively use the
results of \cite{Cvetic:2000db} or \cite{Herzog:2001xk}).
To re-write the K\"ahler form and the metric in the standard form one
starts by introducing left-invariant one forms $\{h_i, \tilde{h}_i\}$
for $(i=1,2,3)$ on the group $SU(2) \times SU(2)$. In terms of the
1-forms $g_i$'s we have \cite{Papadopoulos:2000gj}:

\begin{eqnarray}
\left(
\begin{array}{c}
h_1 \\ h_2
\end{array}
\right)
&=&
\left(
\begin{array}{cc}
   -\cos \frac{\psi}{2}  & -\sin \frac{\psi}{2} \\
   -\sin \frac{\psi}{2}  & \cos \frac{\psi}{2}  
\end{array}
\right)
\left(
\begin{array}{c}
  \frac{1}{\sqrt{2}} (g_1 + g_3) \\ 
  \frac{1}{\sqrt{2}} (g_2 + g_4) 
\end{array}
\right)
 \nonumber \\
\left(
\begin{array}{c}
\tilde{h}_1 \\ \tilde{h}_2
\end{array}
\right)
&=&
\left(
\begin{array}{cc}
   \cos \frac{\psi}{2}  & \sin  \frac{\psi}{2} \\
   -\sin \frac{\psi}{2} &  \cos   \frac{\psi}{2}  
\end{array}
\right)
\left(
\begin{array}{c}
  \frac{1}{\sqrt{2}} (g_3 - g_1) \\ 
  \frac{1}{\sqrt{2}} (g_4 - g_2) 
\end{array}
\right).
\nonumber \\
\textrm{and}  \qquad &&
h_3 + \tilde{h}_3 = g_5.
\end{eqnarray}
Then the K\"ahler 2-form $\Omega$ and the metric will be given by:

\begin{equation}
\Omega = E_1 \wedge E_2  + E_3 \wedge E_4 + E_5 \wedge E_6
\quad \textrm{and} \quad
ds^2_6 = E_1^2 +E_2^2 +E_3^2 +E_4^2 +E_5^2 +E_6^2,
\end{equation}
where

\begin{eqnarray}
E_1 = A(\tau) \left( \alpha(\tau) h_1 - \beta(\tau) \tilde{h}_1 \right)  
&& \qquad
E_2 = A(\tau) \left( \alpha(\tau) h_2 + \beta(\tau) \tilde{h}_2 \right)  
\nonumber \\
E_3 = A(\tau) \left( -\beta(\tau) h_1 + \alpha(\tau) \tilde{h}_1 \right)  
&& \qquad
E_2 = A(\tau) \left( \beta(\tau) h_2 + \alpha(\tau) \tilde{h}_2 \right)  
\nonumber \\
E_5 = B(\tau) d \tau
&& \qquad
E_6 = B(\tau) (h_3 + \tilde{h}_3)
\end{eqnarray}
with

\begin{eqnarray}
A(\tau) &=& \half \epsilon^{2/3} \left( \coth (\tau) (\sinh (2 \tau) - 2 \tau)^{1/3} \right)^{1/2}  
\nonumber \\
B(\tau) &=& \frac{1}{\sqrt{3}} \epsilon^{2/3} \sinh (\tau) (\sinh (2 \tau) - 2 \tau)^{-1/3}  
\nonumber \\
\alpha(\tau) &=& \left( \half \left( 1 + \tanh (\tau) \right) \right)^{1/2}
\qquad \textrm{and} \qquad
\beta(\tau) = \left( \half \left( 1 - \tanh (\tau) \right) \right)^{1/2}.
\end{eqnarray}
The complex structure is defined in terms of the 1-forms $E_i$'s as follows:

\begin{eqnarray}
& J(E_1) = E_2, \quad J(E_2) = -E_1, \quad 
  J(E_3) = E_4, \quad J(E_4) = -E_3, \quad  &
\nonumber\\
& J(E_5) = E_6, \quad J(E_6) = -E_5. \quad  &
\end{eqnarray}
The integrability of this structure can be verified by a straightforward
computation \cite{Papadopoulos:2000gj}.
We are mainly interested in the complex $(0,3)$ form:

\begin{equation}
\eta_{(0,3)} = \frac{2 \sqrt{3}}{\epsilon^2} \mathbf{M} (E_1 - i E_2) \wedge (E_3 - i E_4) \wedge (E_5 - i E_6).
\end{equation}
On plugging the expressions for $E_i$'s  a somewhat lengthy
calculation leads to:

\begin{eqnarray}
\textrm{Re}(\eta_{(0,3)}) &=& \frac{2 \sqrt{3}}{\epsilon^2} \mathbf{M}
                            \left[
                              \left( E_1 \wedge E_3 - E_2 \wedge E_4 \right) \wedge E_5
                            - \left( E_2 \wedge E_3 + E_1 \wedge E_4 \right) \wedge E_6
                            \right]
\nonumber \\
    &=&  \frac{2 \sqrt{3}}{\epsilon^2} \mathbf{M} A^2(\tau) B(\tau)
         \Big[ - (\alpha^2(\tau) - \beta^2(\tau) )
                d\tau \wedge \left( g^1 \wedge g^3 + g^2 \wedge g^4 \right) +
\nonumber \\
    & &       + g_5 \wedge \left( (-1+2\alpha(\tau) \beta(\tau)) g_1 \wedge g_2 
                             + (1+2\alpha(\tau) \beta(\tau)) g_3 \wedge g_4 \right)
          \Big]
\nonumber \\
    &=&  \mathbf{M} \bigg[ -\half \sinh(\tau) d\tau \wedge \left( g^1 \wedge g^3 + g^2 \wedge g^4 \right) +
\nonumber \\
    & &       + \half (1-\cosh(\tau)) g_5 \wedge g_1 \wedge g_2  + \half (1+ \cosh(\tau)) g_5 \wedge g_3 \wedge g_4 
         \bigg]
\end{eqnarray}
and

\begin{eqnarray}
\textrm{Im}(\eta_{(0,3)}) &=& \frac{2 \sqrt{3}}{\epsilon^2} \mathbf{M}
                            \left[
                             - \left( E_2 \wedge E_3 + E_1 \wedge E_4 \right) \wedge E_5
                            + \left( - E_1 \wedge E_3 + E_2 \wedge E_4 \right) \wedge E_6
                            \right]
\nonumber \\
    &=&  \frac{2 \sqrt{3}}{\epsilon^2} \mathbf{M} A^2(\tau) B(\tau)
         \Big[ (\alpha^2(\tau) - \beta^2(\tau) )
                g_5 \wedge \left( g^1 \wedge g^3 + g^2 \wedge g^4 \right) +
\nonumber \\
    & &       + d \tau \wedge \left( (-1+2\alpha(\tau) \beta(\tau)) g_1 \wedge g_2 
                             + (1+2\alpha(\tau) \beta(\tau)) g_3 \wedge g_4 \right)
          \Big]
\nonumber \\
    &=&  \mathbf{M} \bigg[ \half \sinh(\tau) g_5 \wedge \left( g^1 \wedge g^3 + g^2 \wedge g^4 \right) +
\nonumber \\
    & &       + \half (1-\cosh(\tau)) d \tau \wedge g_1 \wedge g_2  +
                 \half (1+ \cosh(\tau)) d \tau \wedge g_3 \wedge g_4 
         \bigg].
\end{eqnarray}
This means that:

\begin{eqnarray}  
F(\tau) &=&  \half ( 1 - \cosh (\tau) )
\nonumber \\
f(\tau) &=&  \half ( \sinh (\tau) - \tau )
\nonumber \\           
k(\tau) &=&  \half ( \sinh (\tau) + \tau )
\end{eqnarray}
which matches  (\ref{eq:fkFsol}) for $C_1=-\half$ and $C_2=C_3=0$.

\bibliography{KSll}

\end{document}